\documentclass[sigconf]{acmart}

\setcopyright{none}
\renewcommand\footnotetextcopyrightpermission[1]{}

\usepackage{amsmath,amsfonts,mathtools}
\usepackage{algorithmic}
\usepackage{graphicx}
\usepackage{textcomp}
\usepackage{xcolor}
\usepackage{braket}
\usepackage{booktabs}
\usepackage{listings}
\usepackage[T1]{fontenc}
\usepackage{hyperref}
\usepackage[capitalise]{cleveref}
\usepackage{dsfont}
\usepackage{tikz}
\usetikzlibrary{decorations.pathreplacing,decorations.pathmorphing}
\usepackage{multirow}
\usepackage{array}

\usepackage{subfig} %to combine chemistry figures

\newcommand{\qmpi}[1]{\texttt{QMPI\_#1}}
\newcommand{\qmpiarg}[1]{\texttt{#1}}

\definecolor{dgreen}{rgb}{0.05,0.05,0.75}
\lstdefinestyle{cppstyle}{
	language={c++},
	basicstyle=\footnotesize\ttfamily, % Global Code Style
	captionpos=b, % Position of the Caption (t for top, b for bottom)
	extendedchars=true, % Allows 256 instead of 128 ASCII characters
	tabsize=2, % number of spaces indented when discovering a tab 
	columns=fixed, % make all characters equal width
	keepspaces=true, % does not ignore spaces to fit width, convert tabs to spaces
	showstringspaces=false, % lets spaces in strings appear as real spaces
	breaklines=true, % wrap lines if they don't fit
	framesep=4pt, % quarter circle size of the round corners
	numbers=none,%left, % show line numbers at the left
	numberstyle=\tiny\textrm, % style of the line numbers
	commentstyle=\color{dgreen}, % style of comments
	keywordstyle=\bfseries, % style of keywords
	backgroundcolor =\color{white},
	numbersep=5pt,
	xleftmargin=3ex
}

\graphicspath{{fig}}

\newcommand\thefont{\expandafter\string\the\font}

\begin{document}

\title{Distributed Quantum Computing with QMPI}

\author{Thomas H\"aner}
\affiliation{
	\institution{Microsoft Quantum}
	\country{Switzerland}
}

\author{Damian S. Steiger}
\affiliation{
	\institution{Microsoft Quantum}
	\country{Switzerland}
}

\author{Torsten Hoefler}
\affiliation{
	\institution{ETH Z\"urich}
	\country{Switzerland}
}

\author{Matthias Troyer}
\affiliation{
	\institution{Microsoft Quantum}
	\country{USA}
}

\begin{abstract}
	Practical applications of quantum computers require millions of physical qubits and it will be challenging for individual quantum processors to reach such qubit numbers. It is therefore timely to investigate the resource requirements of quantum algorithms in a distributed setting, where multiple quantum processors are interconnected by a coherent network. We introduce an extension of the Message Passing Interface (MPI) to enable high-performance implementations of distributed quantum algorithms. In turn, these implementations can be used for testing, debugging, and resource estimation.
	In addition to a prototype implementation of quantum MPI, we present a performance model for distributed quantum computing, SENDQ. The model is inspired by the classical LogP model, making it useful to inform algorithmic decisions when programming distributed quantum computers. Specifically, we consider several optimizations of two quantum algorithms for problems in physics and chemistry, and we detail their effects on performance in the SENDQ model.
\end{abstract}

\maketitle
\pagestyle{plain}

\section{Introduction}
Quantum computing promises to solve certain computational tasks exponentially faster than classical computers, with application domains ranging from cryptography~\cite{shor1994algorithms} to chemistry and material science \cite{RevModPhys.92.015003}.
A host of case studies investigate the minimal requirements for quantum computers to yield a practical advantage~\cite{lee2020even,gidney2019factor,von2020quantum,reiher2017elucidating,scherer2017concrete}. While the resource requirements seem generally feasible, e.g., for applications in computational catalysis~\cite{von2020quantum} and for breaking RSA~\cite{gidney2019factor}, such applications require millions of physical qubits.
Given current projections~\cite{IBMroadmap,googleroadmap}, it will be challenging for individual quantum processors to achieve such qubit numbers. Consequently, these applications may require that computations are distributed across multiple entangled quantum processors.

In a distributed setting, multiple smaller quantum chips are connected coherently, allowing for inter-node communication of quantum information.
For example, IBM's road\-map for large scale devices containing more than 1 million qubit is planned as a set of individual systems with quantum interconnects linking many dilution refrigerators \cite{IBMroadmap} and Google Quantum AI have communicated plans to connect 100 tiles of 10,000 physical qubits each to reach a million physical qubits~\cite{googleroadmap}. For an overview of possible paths to distributed quantum computing, we refer the reader to reviews on the topic~\cite{PRXQuantum.2.017002, van2016path}.

\textbf{Related Work.} There exists a host of software frameworks, programming languages, and compilers for quantum computing~\cite{svore2018q,bichsel2020silq,steiger2018projectq,aleksandrowicz2019qiskit,green2013quipper,javadiabhari2015scaffcc}. However, to the best of our knowledge, no existing framework for quantum computing allows for development of distributed algorithms.

Moreover, progress has been made on simulations and applications of a quantum internet~\cite{dahlberg2018simulaqron, dahlberg2019link,wehner2018quantum}. Yet, as with today's classical internet applications, these works do not aim to provide a framework for high-performance distributed quantum computing. Instead, typical use cases of a quantum internet are secure communication, clock synchronization and leader election~\cite{wehner2018quantum, pirandola2020advances}.

There exists a large body of theoretical work to estimate the resource requirements of non-local operations \cite{PhysRevA.64.032302, PhysRevA.62.052317}, of distributed quantum algorithm primitives such as distributed arithmetic~\cite{meter2008arithmetic} and the quantum Fourier transform~\cite{yimsiriwattana2004generalized}, and of entire applications in cryptography \cite{yimsiriwattana2004distributed, meter2006architecture, gidney2019factor} and computational chemistry \cite{diadamo2021distributed}.

Finally, there is related work on theoretical models of distributed quantum computing. Beals \emph{et al.}~\cite{beals2013efficient} introduce the Q PRAM model, the shared quantum memory equivalent of the PRAM model with global load/store access as a model for distributed quantum computing. They analyze several quantum algorithms in the Q PRAM model. In contrast to our work, however, their focus is on asymptotic runtimes, and not on performance.

While we consider systems where each node has a sufficient number of physical qubits to encode several logical qubits, we note that there are alternative approaches. For example, \citet{PhysRevX.4.041041} propose a protocol for distributed quantum computing in which small cells of only 5 to 50 physical qubits are connected. In this setting, even a single logical qubit is spread over different nodes, and the distribution is a hardware implementation detail not exposed to the user.

\textbf{Contributions.} In order to bridge the gap between theoretical distributed quantum algorithms and software frameworks for quantum computing, we propose Quantum MPI (QMPI), an extension of the classical MPI standard~\cite{walker1996mpi} to quantum computing. The focus of QMPI is to provide the primitives that are necessary to implement high-performance distributed quantum algorithms. To reason about the performance of distributed quantum algorithms, we develop the SENDQ model and we present examples that illustrate how this model may be leveraged to inform algorithmic decisions.

Specifically, our contributions are as follows:
\begin{itemize}
	\item We define Quantum MPI (QMPI) as an extension to classical MPI. QMPI supports all classical MPI functionality on computational basis states (including their inverses to enable reversibility) as well as general-purpose point-to-point and collective functions that \textit{entangle} and \textit{move} qubits between nodes.
	\item We present a quantum communication model (SENDQ) that is inspired by the classical LogP model~\cite{culler1993logp} to model the performance of distributed quantum algorithms and foster algorithmic optimizations.
 	\item We implement quantum-specific optimizations, such as using asynchronously pre-established entangled quantum states to optimize point-to-point and collective quantum communication with zero quantum communication depth and purely classical communication.
	\item We discuss potential applications of quantum MPI to problems from physics and chemistry, and we show how such applications can be optimized using the SENDQ model.
\end{itemize}

QMPI adds support for quantum message passing to existing quantum programming languages, thus enabling programmers to implement distributed quantum algorithms. In combination with SENDQ, the resulting implementations can be used to investigate typical workloads at application scale. The results of such investigations are crucial for making informed architectural decisions along the road to practical distributed quantum computing.

\section{Quantum Computing}\label{sec:qcintro}

This section serves as a brief introduction to quantum computing and our notation. For a more detailed treatment of this subject, we refer the reader to the textbook by Nielsen and Chuang~\cite{nielsen2002quantum}.

\textbf{Quantum States and Dirac Notation.} A quantum computer consists of multiple quantum bits (qubits) whose quantum state may be represented as a complex superposition over all classical bitstrings. Specifically, the state $\ket\psi$ (``ket $\psi$'') of an $n$-qubit quantum computer may be written as
\[
	\ket\psi = \sum_{i=0}^{2^n-1} \alpha_i\ket{i_{n-1}\cdots i_0},
\]
where $i_k$ denotes the $k$th bit of the integer $i$, and $\alpha_i\in\mathbb C$ such that $\sum_i |\alpha_i|^2=1$. When measuring all qubits at once, the probability of observing the integer $i$ is given by $|\alpha_i|^2$. This also explains the normalization condition, since the probability of observing any integer should be equal to $1$. The ``ket'' notation $\ket{\cdot}$ denotes column vectors, whereas row vectors are denoted by ``bra'': $\bra{\cdot}$. Therefore, the dot-product between two state vectors $\ket\psi,\ket\phi$ can be written as $\braket{\psi|\phi}$ and the projector onto $\ket\psi$ is $P_{\ket\psi}:=\ket\psi\bra\psi$. By identifying each of the computational basis states $\ket{i_{n-1}\cdots i_0}$ with the corresponding standard basis vector $e_i\in\mathbb C^{2^n}$, i.e., $(e_i)_j=\delta_{ij}$ where $\delta_{ij}$ denotes the Kronecker delta, $\ket\psi$ can be written as a column vector with entries $(\ket\psi)_i=\alpha_i$.

\textbf{Quantum Gates.} A computation can be performed by applying a sequence of quantum instructions to the qubits. Such quantum instructions consist of a list of qubit indices that the instruction acts upon, and a so-called quantum gate, akin to classical gates such as AND, XOR, etc.. In the same way that quantum states can be represented as column vectors, quantum gates may be represented as (complex) unitary matrices $U$ of dimension $2^n\times 2^n$. A matrix is said to be unitary iff $U^\dagger U=UU^\dagger = \mathds 1$, where $U^\dagger$ denotes the Hermitian conjugate of $U$. Then, matrix-vector multiplication models the application of a quantum gate.

We will be using the following gates in this paper. The Hadamard gate $H=\frac1{\sqrt2}\begin{psmallmatrix} 1&1\\1&-1\end{psmallmatrix}$, the $S$ gate $S=\begin{psmallmatrix} 1&0\\0&i\end{psmallmatrix}$, the Pauli gates $X=\begin{psmallmatrix}
	0&1\\1&0
\end{psmallmatrix},
Y=\begin{psmallmatrix}
	0&-i\\i&0
\end{psmallmatrix},
Z=\begin{psmallmatrix}
	1&0\\0&-1
\end{psmallmatrix}$,
the controlled Pauli gates, e.g., controlled X (or controlled NOT, CNOT)
\[
	CX=\ket0\bra0\otimes\mathds 1_2+\ket1\bra1\otimes X=\begin{psmallmatrix} 1&0&0&0\\0&1&0&0\\0&0&0&1\\0&0&1&0\end{psmallmatrix},
\]
(where $\otimes$ denotes the Kronecker product), and Pauli rotation gates
\[
	R_{P}(\theta)=e^{-0.5i\theta P},
\]
where $P$ is a single-qubit Pauli gate $P\in\{X,Y,Z\}$.

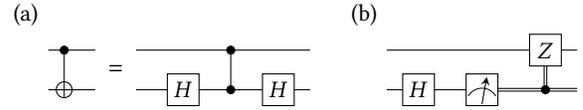
\begin{figure}
	\centering
	\begin{tikzpicture}[scale=1.000000,x=1pt,y=1pt]
\node at (-0.3cm,1cm) {(a)};
\filldraw[color=white] (0.000000, -7.500000) rectangle (99.000000, 22.500000);
% Drawing wires
% Line 1: q1 W {} {}
\draw[color=black] (0.000000,15.000000) -- (99.000000,15.000000);
\draw[color=black] (0.000000,15.000000) node[left] {${}$};
% Line 2: q2 W {} {}
\draw[color=black] (0.000000,0.000000) -- (99.000000,0.000000);
\draw[color=black] (0.000000,0.000000) node[left] {${}$};
% Done with wires; drawing gates
% Line 4: q1 +q2 length=0
\draw (6.000000,15.000000) -- (6.000000,0.000000);
\filldraw (6.000000, 15.000000) circle(1.500000pt);
\begin{scope}
\draw[fill=white] (6.000000, 0.000000) circle(3.000000pt);
\clip (6.000000, 0.000000) circle(3.000000pt);
\draw (3.000000, 0.000000) -- (9.000000, 0.000000);
\draw (6.000000, -3.000000) -- (6.000000, 3.000000);
\end{scope}
% Line 6: =
\draw[fill=white,color=white] (18.000000, -6.000000) rectangle (33.000000, 21.000000);
\draw (25.500000, 7.500000) node {$=$};
% Line 8: q2 H
\begin{scope}
\draw[fill=white] (51.000000, -0.000000) +(-45.000000:8.485281pt and 8.485281pt) -- +(45.000000:8.485281pt and 8.485281pt) -- +(135.000000:8.485281pt and 8.485281pt) -- +(225.000000:8.485281pt and 8.485281pt) -- cycle;
\clip (51.000000, -0.000000) +(-45.000000:8.485281pt and 8.485281pt) -- +(45.000000:8.485281pt and 8.485281pt) -- +(135.000000:8.485281pt and 8.485281pt) -- +(225.000000:8.485281pt and 8.485281pt) -- cycle;
\draw (51.000000, -0.000000) node {$H$};
\end{scope}
% Line 9: q1 q2 length=0
\draw (69.000000,15.000000) -- (69.000000,0.000000);
\filldraw (69.000000, 15.000000) circle(1.500000pt);
\filldraw (69.000000, 0.000000) circle(1.500000pt);
% Line 10: q2 H
\begin{scope}
\draw[fill=white] (87.000000, -0.000000) +(-45.000000:8.485281pt and 8.485281pt) -- +(45.000000:8.485281pt and 8.485281pt) -- +(135.000000:8.485281pt and 8.485281pt) -- +(225.000000:8.485281pt and 8.485281pt) -- cycle;
\clip (87.000000, -0.000000) +(-45.000000:8.485281pt and 8.485281pt) -- +(45.000000:8.485281pt and 8.485281pt) -- +(135.000000:8.485281pt and 8.485281pt) -- +(225.000000:8.485281pt and 8.485281pt) -- cycle;
\draw (87.000000, -0.000000) node {$H$};
\end{scope}
% Done with gates; drawing ending labels
\draw[color=black] (99.000000,15.000000) node[right] {${}$};
\draw[color=black] (99.000000,0.000000) node[right] {${}$};
% Done with ending labels; drawing cut lines and comments
% Done with comments

\node at (4.2cm,1cm) {(b)};
\begin{scope}[xshift=4.5cm]
\filldraw[color=white] (0.000000, -7.500000) rectangle (72.000000, 22.500000);
% Drawing wires
% Line 1: q1 W {} {}
\draw[color=black] (0.000000,15.000000) -- (72.000000,15.000000);
\draw[color=black] (0.000000,15.000000) node[left] {${}$};
% Line 2: q2 W {} {}
\draw[color=black] (0.000000,0.000000) -- (36.000000,0.000000);
\draw[color=black] (36.000000,-0.500000) -- (72.000000,-0.500000);
\draw[color=black] (36.000000,0.500000) -- (72.000000,0.500000);
\draw[color=black] (0.000000,0.000000) node[left] {${}$};
% Done with wires; drawing gates
% Line 4: q2 H
\begin{scope}
\draw[fill=white] (12.000000, -0.000000) +(-45.000000:8.485281pt and 8.485281pt) -- +(45.000000:8.485281pt and 8.485281pt) -- +(135.000000:8.485281pt and 8.485281pt) -- +(225.000000:8.485281pt and 8.485281pt) -- cycle;
\clip (12.000000, -0.000000) +(-45.000000:8.485281pt and 8.485281pt) -- +(45.000000:8.485281pt and 8.485281pt) -- +(135.000000:8.485281pt and 8.485281pt) -- +(225.000000:8.485281pt and 8.485281pt) -- cycle;
\draw (12.000000, -0.000000) node {$H$};
\end{scope}
% Line 5: q2 M
\draw[fill=white] (30.000000, -6.000000) rectangle (42.000000, 6.000000);
\draw[very thin] (36.000000, 0.600000) arc (90:150:6.000000pt);
\draw[very thin] (36.000000, 0.600000) arc (90:30:6.000000pt);
\draw[->,>=stealth] (36.000000, -5.400000) -- +(80:10.392305pt);
% Line 6: q1 Z q2
\draw (59.500000,15.000000) -- (59.500000,0.000000);
\draw (60.500000,15.000000) -- (60.500000,0.000000);
\begin{scope}
\draw[fill=white] (60.000000, 15.000000) +(-45.000000:8.485281pt and 8.485281pt) -- +(45.000000:8.485281pt and 8.485281pt) -- +(135.000000:8.485281pt and 8.485281pt) -- +(225.000000:8.485281pt and 8.485281pt) -- cycle;
\clip (60.000000, 15.000000) +(-45.000000:8.485281pt and 8.485281pt) -- +(45.000000:8.485281pt and 8.485281pt) -- +(135.000000:8.485281pt and 8.485281pt) -- +(225.000000:8.485281pt and 8.485281pt) -- cycle;
\draw (60.000000, 15.000000) node {$Z$};
\end{scope}
\filldraw (60.000000, 0.000000) circle(1.500000pt);
% Done with gates; drawing ending labels
\draw[color=black] (72.000000,15.000000) node[right] {${}$};
\draw[color=black] (72.000000,0.000000) node[right] {${}$};
% Done with ending labels; drawing cut lines and comments
% Done with comments
\end{scope}
\end{tikzpicture}
	\caption{(a) A CNOT may be written in terms of a CZ using Hadamard gates. (b) If the target qubit is known to be reset to $\ket0$ by the CNOT, then this reset may be implemented more efficiently using only single-qubit gates and classical control.}
	\label{fig:mcnot}
\end{figure}

\textbf{Quantum Circuits.} To illustrate a sequence of quantum gates acting on qubits, we use circuit diagrams such as the one in \cref{fig:mcnot}. Each horizontal line corresponds to a qubit and boxes/symbols on these lines represent gates, with time advancing from left to right. We use double-lines to denote classical information such as measurement outcomes. Controlled gates (such as CNOT) are drawn with as a filled circle $\bullet$ on the control qubit and a line connecting the control qubit to the target qubit gate. Moreover, the Pauli X gate (or NOT gate) is drawn as a $\oplus$-symbol, since it corresponds to addition modulo two, and the controlled Z gate is sometimes drawn as two $\bullet$-symbols connected by a line (to illustrate its symmetry with respect to control and target qubit).

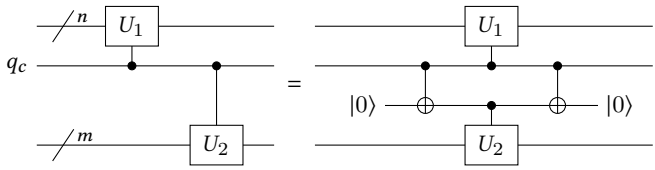
\begin{figure}
	\centering
	\begin{tikzpicture}[scale=1.000000,x=1pt,y=1pt]
\filldraw[color=white] (0.000000, -7.500000) rectangle (233.000000, 52.500000);
% Drawing wires
% Line 2: q0 W
\draw[color=black] (0.000000,45.000000) -- (233.000000,45.000000);
% Line 3: qc W q_c
\draw[color=black] (0.000000,30.000000) -- (233.000000,30.000000);
\draw[color=black] (0.000000,30.000000) node[left] {$q_c$};
% Line 4: qa W |0\rangle |0\rangle
\draw[color=black] (124.500000,15.000000) -- (219.500000,15.000000);
% Line 5: q1 W
\draw[color=black] (0.000000,0.000000) -- (233.000000,0.000000);
% Done with wires; drawing gates
% Line 8: q0 / n
\draw (6.000000, 39.000000) -- (14.000000, 51.000000);
\draw (12.000000, 48.000000) node[right] {$\scriptstyle{n}$};
% Line 9: q1 / m
\draw (6.000000, -6.000000) -- (14.000000, 6.000000);
\draw (12.000000, 3.000000) node[right] {$\scriptstyle{m}$};
% Line 11: q0 G $U_1$ qc width=20 height=15
\draw (36.000000,45.000000) -- (36.000000,30.000000);
\begin{scope}
\draw[fill=white] (36.000000, 45.000000) +(-45.000000:14.142136pt and 10.606602pt) -- +(45.000000:14.142136pt and 10.606602pt) -- +(135.000000:14.142136pt and 10.606602pt) -- +(225.000000:14.142136pt and 10.606602pt) -- cycle;
\clip (36.000000, 45.000000) +(-45.000000:14.142136pt and 10.606602pt) -- +(45.000000:14.142136pt and 10.606602pt) -- +(135.000000:14.142136pt and 10.606602pt) -- +(225.000000:14.142136pt and 10.606602pt) -- cycle;
\draw (36.000000, 45.000000) node {$U_1$};
\end{scope}
\filldraw (36.000000, 30.000000) circle(1.500000pt);
% Line 12: q1 G $U_2$ qc width=20 height=15
\draw (68.000000,30.000000) -- (68.000000,0.000000);
\begin{scope}
\draw[fill=white] (68.000000, -0.000000) +(-45.000000:14.142136pt and 10.606602pt) -- +(45.000000:14.142136pt and 10.606602pt) -- +(135.000000:14.142136pt and 10.606602pt) -- +(225.000000:14.142136pt and 10.606602pt) -- cycle;
\clip (68.000000, -0.000000) +(-45.000000:14.142136pt and 10.606602pt) -- +(45.000000:14.142136pt and 10.606602pt) -- +(135.000000:14.142136pt and 10.606602pt) -- +(225.000000:14.142136pt and 10.606602pt) -- cycle;
\draw (68.000000, -0.000000) node {$U_2$};
\end{scope}
\filldraw (68.000000, 30.000000) circle(1.500000pt);
% Line 15: =
\draw[fill=white,color=white] (90.000000, -6.000000) rectangle (105.000000, 51.000000);
\draw (97.500000, 22.500000) node {$=$};
% Line 17: qa START
\draw[color=black] (132.000000,15.000000) node[fill=white,left,minimum height=15.000000pt,minimum width=15.000000pt,inner sep=0pt] {\phantom{$|0\rangle$}};
\draw[color=black] (132.000000,15.000000) node[left] {$|0\rangle$};
% Line 19: +qa qc
\draw (147.000000,30.000000) -- (147.000000,15.000000);
\begin{scope}
\draw[fill=white] (147.000000, 15.000000) circle(3.000000pt);
\clip (147.000000, 15.000000) circle(3.000000pt);
\draw (144.000000, 15.000000) -- (150.000000, 15.000000);
\draw (147.000000, 12.000000) -- (147.000000, 18.000000);
\end{scope}
\filldraw (147.000000, 30.000000) circle(1.500000pt);
% Line 21: q0 G $U_1$ qc width=20 height=15
\draw (172.000000,45.000000) -- (172.000000,30.000000);
\begin{scope}
\draw[fill=white] (172.000000, 45.000000) +(-45.000000:14.142136pt and 10.606602pt) -- +(45.000000:14.142136pt and 10.606602pt) -- +(135.000000:14.142136pt and 10.606602pt) -- +(225.000000:14.142136pt and 10.606602pt) -- cycle;
\clip (172.000000, 45.000000) +(-45.000000:14.142136pt and 10.606602pt) -- +(45.000000:14.142136pt and 10.606602pt) -- +(135.000000:14.142136pt and 10.606602pt) -- +(225.000000:14.142136pt and 10.606602pt) -- cycle;
\draw (172.000000, 45.000000) node {$U_1$};
\end{scope}
\filldraw (172.000000, 30.000000) circle(1.500000pt);
% Line 22: q1 G $U_2$ qa width=20 height=15
\draw (172.000000,15.000000) -- (172.000000,0.000000);
\begin{scope}
\draw[fill=white] (172.000000, -0.000000) +(-45.000000:14.142136pt and 10.606602pt) -- +(45.000000:14.142136pt and 10.606602pt) -- +(135.000000:14.142136pt and 10.606602pt) -- +(225.000000:14.142136pt and 10.606602pt) -- cycle;
\clip (172.000000, -0.000000) +(-45.000000:14.142136pt and 10.606602pt) -- +(45.000000:14.142136pt and 10.606602pt) -- +(135.000000:14.142136pt and 10.606602pt) -- +(225.000000:14.142136pt and 10.606602pt) -- cycle;
\draw (172.000000, -0.000000) node {$U_2$};
\end{scope}
\filldraw (172.000000, 15.000000) circle(1.500000pt);
% Line 24: +qa qc
\draw (197.000000,30.000000) -- (197.000000,15.000000);
\begin{scope}
\draw[fill=white] (197.000000, 15.000000) circle(3.000000pt);
\clip (197.000000, 15.000000) circle(3.000000pt);
\draw (194.000000, 15.000000) -- (200.000000, 15.000000);
\draw (197.000000, 12.000000) -- (197.000000, 18.000000);
\end{scope}
\filldraw (197.000000, 30.000000) circle(1.500000pt);
% Line 26: qa END
\draw[color=black] (212.000000,15.000000) node[fill=white,right,minimum height=15.000000pt,minimum width=15.000000pt,inner sep=0pt] {\phantom{$|0\rangle$}};
\draw[color=black] (212.000000,15.000000) node[right] {$|0\rangle$};
% Line 28: q0 TOUCH
% Done with gates; drawing ending labels
% Done with ending labels; drawing cut lines and comments
% Done with comments
\end{tikzpicture}
	\caption{Fanout of control qubit $q_c$ to apply the controlled gates $U_1$ and $U_2$ in parallel.}
	\label{fig:fanout_simple}
\end{figure}

We will be using a circuit primitive called \emph{fanout}, which can be viewed as copying in a quantum superposition. Let us assume a  qubit $q_c=\alpha \ket{0} + \beta\ket{1}$  in a superposition  of classical values 0 and 1. Fanout adds auxiliary qubits and transforms the state  to $\alpha \ket{0...0} + \beta\ket{1...1}$., thus it is now in a superposition of multiple copies of the classical values 0 and 1. Note that this is not the same as cloning the qubit.
One application of fanout is to to parallelize computations \cite{hoyer2005quantum} by fanning out control qubits so that gates can be executed in parallel even when they are executed conditionally on the same control qubit(s). Gates that are controlled on $q_c$ and that act on distinct qubits may now be applied in parallel by choosing qubit $q_c$ or any of the auxiliary qubits as a control qubit. After completion of the controlled gates, all auxiliary qubits need to reverted back to 0 by reverse fanout. Locally, this is again done by a simple CNOT gate, see \cref{fig:fanout_simple}.

Fanout is just one example of classical computation applied to a superpositions of states. More generally, any reversible classical computation can be applied in superposition, and this includes many MPI operations to be discussed later, including reductions.

\textbf{EPR Pairs and Quantum Teleportation.} When distributing a quantum algorithm to multiple nodes, some multi-qubit gates act on qubits that are located on different nodes. This situation can be resolved through either gate or qubit teleportation. We will briefly review the latter and we refer the reader to the work by \citet{yimsiriwattana2004generalized} for a more detailed discussion.

The fundamental resource to enable communication of quantum data are Einstein-Podolsky-Rosen (EPR) pairs~\cite{PhysRev.47.777}, which consist of two qubits in the state
\[
	\frac 1{\sqrt2}(\ket{00}+\ket{11}).
\]
In a circuit diagram, we depict EPR pairs as two filled circles that are interconnected with a serpentine line \tikz[x=1pt,y=1pt]{
	\node[circle,minimum size=3pt,inner sep=0pt,outer sep=0pt,fill=black,draw] (1) at (0,0) {};
	\node[circle,minimum size=3pt,inner sep=0pt,outer sep=0pt,fill=black,draw] (2) at (15,0) {};
	\draw[decorate,decoration={snake}] (1) -- (2);
}.

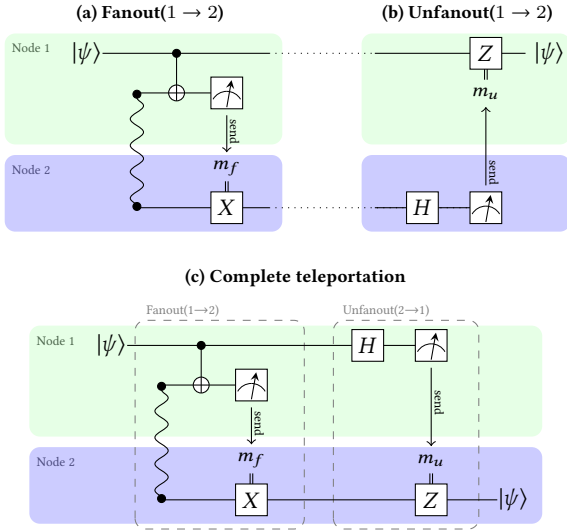
\begin{figure}
	\centering
	%! \usetikzlibrary{decorations.pathreplacing,decorations.pathmorphing}
\begin{tikzpicture}[scale=1.000000,x=1pt,y=1pt]
%\fill[color=green] (0,120) rectangle (120, -120);

\fill[color=green!10!white,rounded corners] (-70.000000, -4.500000) rectangle (35.000000, 37.500000);
\fill[color=green!10!white,rounded corners] (65.000000, -4.500000) rectangle (144.000000, 37.500000);

\fill[color=blue!20!white,rounded corners] (-70.000000, -8.500000) rectangle (35.000000, -37.500000);
\fill[color=blue!20!white,rounded corners] (65.000000, -8.500000) rectangle (144.000000, -37.500000);

\begin{scope}[xshift=-20]
\node at (-40,32) {\textcolor[rgb]{0.3,0.4,0.3}{\tiny Node 1}};
\node at (-40,-14) {\textcolor[rgb]{0.3,0.3,0.5}{\tiny Node 2}};

%\draw[color=black!50!white,dashed,rounded corners] (-10.000000, -39.5) rectangle (54.000000, 39.500000);
%\node at (8,43) {\textcolor{gray}{\tiny Fanout(1$\rightarrow$2)}};

%\draw[color=black!50!white,dashed,rounded corners] (65.000000, -39.5) rectangle (120.000000, 39.500000);
%\node at (85,43) {\textcolor{gray}{\tiny Unfanout(1$\rightarrow$2)}};

% Drawing wires
% Line 1: a W |\psi\rangle
\draw[color=black] (-13.000000,30.000000) -- (50.000000,30.000000);
\draw[color=black,dotted] (50.000000,30.000000) -- (90.000000,30.000000);
\draw[color=black] (90.000000,30.000000) -- (147.000000,30.000000);

%\draw[color=black] (54.000000,29.500000) -- (68.000000,29.500000);
%\draw[color=black] (54.000000,30.500000) -- (68.000000,30.500000);
\draw[color=black] (-10.000000,30.000000) node[left] {$|\psi\rangle$};
\draw[color=black] (147.000000,30.000000) node[right] {$|\psi\rangle$};

% Line 2: b c W \frac1{\sqrt2}(|00\rangle+|11\rangle)<
\draw[color=black] (0.000000,15.000000) -- (28.000000,15.000000);
%\draw[color=black] (54.000000,14.500000) -- (68.000000,14.500000);
%\draw[color=black] (54.000000,15.500000) -- (68.000000,15.500000);

% message:
%\draw[decorate,decoration={brace,amplitude = 3.000000pt,mirror}] (68.000000,11.500000) -- (68.000000,33.500000);
%\node (meas) at (78,22.5) {$m$};
%\node (recv) at (90,-10) {};
%\path[->] (meas.east) edge[bend left=30] node[pos=0.5,right] {\footnotesize send} (recv.north);

%   Deferring wire label at (0.000000,15.000000)
% Line 2: b c W \frac1{\sqrt2}(|00\rangle+|11\rangle)<

%\filldraw[color=white,fill=white] (0.000000,-3.750000) rectangle (-4.000000,18.750000);
%\draw[decorate,decoration={brace,amplitude = 4.000000pt}] (0.000000,-3.750000) -- (0.000000,18.750000);

% draw EPR pair
\begin{scope}[yshift=-0.5cm]
%\node[color=black,fill=white,circle] at (-5.000000,7.00000) {\rotatebox{90}{\tiny EPR}};
\draw[decorate,decoration={snake}] (0.000000,-13.750000) -- (0.000000,28.750000);
\filldraw (0.000000,-13.750000) circle(1.500000pt);
\filldraw (0.000000,28.750000) circle(1.500000pt);
\end{scope}
% Done with wires; drawing gates
% Line 4: a +b
\begin{scope}[xshift=6]
	\draw (9.000000,30.000000) -- (9.000000,15.000000);
	\filldraw (9.000000, 30.000000) circle(1.500000pt);
	\draw[fill=white] (9.000000, 15.000000) circle(3.000000pt);
	\clip (9.000000, 15.000000) circle(3.000000pt);
	\draw (6.000000, 15.000000) -- (12.000000, 15.000000);
	\draw (9.000000, 12.000000) -- (9.000000, 18.000000);
\end{scope}

\begin{scope}[xshift=-20]
	\draw[fill=white] (48.000000, 9.000000) rectangle (60.000000, 21.000000);
	\draw[very thin] (54.000000, 15.600000) arc (90:150:6.000000pt);
	\draw[very thin] (54.000000, 15.600000) arc (90:30:6.000000pt);
	\draw[->,>=stealth] (54.000000, 9.600000) -- +(80:10.392305pt);
\end{scope}

\end{scope}

\begin{scope}[yshift=30,xshift=10]
% measurement lines
\draw (101.500000,-10.000000) -- (101.500000,0.000000);
\draw (102.500000,-10.000000) -- (102.500000,0.000000);
\begin{scope}
	\draw[fill=white] (102.000000, -0.000000) +(-45.000000:8.485281pt and 8.485281pt) -- +(45.000000:8.485281pt and 8.485281pt) -- +(135.000000:8.485281pt and 8.485281pt) -- +(225.000000:8.485281pt and 8.485281pt) -- cycle;
	\clip (102.000000, -0.000000) +(-45.000000:8.485281pt and 8.485281pt) -- +(45.000000:8.485281pt and 8.485281pt) -- +(135.000000:8.485281pt and 8.485281pt) -- +(225.000000:8.485281pt and 8.485281pt) -- cycle;
	\draw (102.000000, -0.000000) node {$Z$};
\end{scope}
\node (m2) at  (102.000000, -15.000000)  {${\scriptstyle m_u}$};
\draw[->] (102,-50)--(m2);
\node at (105,-45) {\rotatebox{-90}{\tiny send}};
\end{scope}

\begin{scope}[yshift=-1cm]
\begin{scope}[xshift=-20]
% draw line for qubit on second node:
\draw[color=black] (0.000000,0.000000) -- (50.000000,0.000000);
\node (n1) at (50,0) {};

% Line 9: c X b:owire
\begin{scope}[xshift=-44]
	\draw (77.500000,10.000000) -- (77.500000,0.000000);
	\draw (78.500000,10.000000) -- (78.500000,0.000000);

	\draw[fill=white] (78.000000, -0.000000) +(-45.000000:8.485281pt and 8.485281pt) -- +(45.000000:8.485281pt and 8.485281pt) -- +(135.000000:8.485281pt and 8.485281pt) -- +(225.000000:8.485281pt and 8.485281pt) -- cycle;
	\clip (78.000000, -0.000000) +(-45.000000:8.485281pt and 8.485281pt) -- +(45.000000:8.485281pt and 8.485281pt) -- +(135.000000:8.485281pt and 8.485281pt) -- +(225.000000:8.485281pt and 8.485281pt) -- cycle;
	\draw (78.000000, -0.000000) node {$X$};
\end{scope}
\node (m1) at (34.000000, 15.000000) {${\scriptstyle m_f}$};
\draw[->] (34,35)--(m1);
% Line 10: c Z b:owire # this is wrong (used to simplify manual editing of tikz)
\end{scope}
%%%%%%%%%%%%
\begin{scope}[xshift=10]
\node (n2) at (60,0) {};
\draw[color=black] (60,0) -- (102.000000,0.000000);
\draw[color=black,dotted] (20,0) -- (100,0);
% Line 6: a b M

\begin{scope}[xshift=48,yshift=-30]
	\draw[fill=white] (30.000000, 30.000000) +(-45.000000:8.485281pt and 8.485281pt) -- +(45.000000:8.485281pt and 8.485281pt) -- +(135.000000:8.485281pt and 8.485281pt) -- +(225.000000:8.485281pt and 8.485281pt) -- cycle;
	\clip (30.000000, 30.000000) +(-45.000000:8.485281pt and 8.485281pt) -- +(45.000000:8.485281pt and 8.485281pt) -- +(135.000000:8.485281pt and 8.485281pt) -- +(225.000000:8.485281pt and 8.485281pt) -- cycle;
	\draw (30.000000, 30.000000) node {$H$};
\end{scope}
\begin{scope}[xshift=48,yshift=-30]
	\draw[fill=white] (48.000000, 24.000000) rectangle (60.000000, 36.000000);
	\draw[very thin] (54.000000, 30.600000) arc (90:150:6.000000pt);
	\draw[very thin] (54.000000, 30.600000) arc (90:30:6.000000pt);
	\draw[->,>=stealth] (54.000000, 24.600000) -- +(80:10.392305pt);
\end{scope}

\node at (7,29) {\rotatebox{-90}{\tiny send}};
% Done with gates; drawing ending labels
%\draw[color=black] (124.000000,0.000000) node[right] {$|\psi\rangle$};
% Done with ending labels; drawing cut lines and comments
% Done with comments
\end{scope}
\end{scope}

\node at (-15,45) {\footnotesize \textbf{(a) Fanout($1\rightarrow 2$)}};
\node at (105,45) {\footnotesize \textbf{(b) Unfanout($1\rightarrow 2$)}};
\node at (40,-55) {\footnotesize \textbf{(c) Complete teleportation}};
\end{tikzpicture}
	%! \usetikzlibrary{decorations.pathreplacing,decorations.pathmorphing}
\begin{tikzpicture}[scale=1.000000,x=1pt,y=1pt]
%\fill[color=green] (0,120) rectangle (120, -120);

\fill[color=green!10!white,rounded corners] (-50.000000, -4.500000) rectangle (144.000000, 37.500000);
\fill[color=blue!20!white,rounded corners] (-50.000000, -8.500000) rectangle (144.000000, -37.500000);

\node at (-40,32) {\textcolor[rgb]{0.3,0.4,0.3}{\tiny Node 1}};
\node at (-40,-14) {\textcolor[rgb]{0.3,0.3,0.5}{\tiny Node 2}};

\draw[color=black!50!white,dashed,rounded corners] (-10.000000, -39.5) rectangle (54.000000, 39.500000);
\node at (8,43) {\textcolor{gray}{\tiny Fanout(1$\rightarrow$2)}};

\draw[color=black!50!white,dashed,rounded corners] (65.000000, -39.5) rectangle (120.000000, 39.500000);
\node at (85,43) {\textcolor{gray}{\tiny Unfanout(2$\rightarrow$1)}};

% Drawing wires
% Line 1: a W |\psi\rangle
\draw[color=black] (-13.000000,30.000000) -- (104.000000,30.000000);
%\draw[color=black] (54.000000,29.500000) -- (68.000000,29.500000);
%\draw[color=black] (54.000000,30.500000) -- (68.000000,30.500000);
\draw[color=black] (-10.000000,30.000000) node[left] {$|\psi\rangle$};
% Line 2: b c W \frac1{\sqrt2}(|00\rangle+|11\rangle)<
\draw[color=black] (0.000000,15.000000) -- (28.000000,15.000000);
%\draw[color=black] (54.000000,14.500000) -- (68.000000,14.500000);
%\draw[color=black] (54.000000,15.500000) -- (68.000000,15.500000);

% message:
%\draw[decorate,decoration={brace,amplitude = 3.000000pt,mirror}] (68.000000,11.500000) -- (68.000000,33.500000);
%\node (meas) at (78,22.5) {$m$};
%\node (recv) at (90,-10) {};
%\path[->] (meas.east) edge[bend left=30] node[pos=0.5,right] {\footnotesize send} (recv.north);

%   Deferring wire label at (0.000000,15.000000)
% Line 2: b c W \frac1{\sqrt2}(|00\rangle+|11\rangle)<

%\filldraw[color=white,fill=white] (0.000000,-3.750000) rectangle (-4.000000,18.750000);
%\draw[decorate,decoration={brace,amplitude = 4.000000pt}] (0.000000,-3.750000) -- (0.000000,18.750000);

% draw EPR pair
\begin{scope}[yshift=-0.5cm]
%\node[color=black,fill=white,circle] at (-5.000000,7.00000) {\rotatebox{90}{\tiny EPR}};
\draw[decorate,decoration={snake}] (0.000000,-13.750000) -- (0.000000,28.750000);
\filldraw (0.000000,-13.750000) circle(1.500000pt);
\filldraw (0.000000,28.750000) circle(1.500000pt);
\end{scope}
% Done with wires; drawing gates
% Line 4: a +b
\begin{scope}[xshift=6]
	\draw (9.000000,30.000000) -- (9.000000,15.000000);
	\filldraw (9.000000, 30.000000) circle(1.500000pt);
	\draw[fill=white] (9.000000, 15.000000) circle(3.000000pt);
	\clip (9.000000, 15.000000) circle(3.000000pt);
	\draw (6.000000, 15.000000) -- (12.000000, 15.000000);
	\draw (9.000000, 12.000000) -- (9.000000, 18.000000);
\end{scope}
% Line 5: a H
\begin{scope}[xshift=48]
	\draw[fill=white] (30.000000, 30.000000) +(-45.000000:8.485281pt and 8.485281pt) -- +(45.000000:8.485281pt and 8.485281pt) -- +(135.000000:8.485281pt and 8.485281pt) -- +(225.000000:8.485281pt and 8.485281pt) -- cycle;
	\clip (30.000000, 30.000000) +(-45.000000:8.485281pt and 8.485281pt) -- +(45.000000:8.485281pt and 8.485281pt) -- +(135.000000:8.485281pt and 8.485281pt) -- +(225.000000:8.485281pt and 8.485281pt) -- cycle;
	\draw (30.000000, 30.000000) node {$H$};
\end{scope}
% Line 6: a b M
\begin{scope}[xshift=48]
\draw[fill=white] (48.000000, 24.000000) rectangle (60.000000, 36.000000);
\draw[very thin] (54.000000, 30.600000) arc (90:150:6.000000pt);
\draw[very thin] (54.000000, 30.600000) arc (90:30:6.000000pt);
\draw[->,>=stealth] (54.000000, 24.600000) -- +(80:10.392305pt);
\end{scope}
\begin{scope}[xshift=-20]
\draw[fill=white] (48.000000, 9.000000) rectangle (60.000000, 21.000000);
\draw[very thin] (54.000000, 15.600000) arc (90:150:6.000000pt);
\draw[very thin] (54.000000, 15.600000) arc (90:30:6.000000pt);
\draw[->,>=stealth] (54.000000, 9.600000) -- +(80:10.392305pt);
\end{scope}

\begin{scope}[yshift=-1cm]
% draw line for qubit on second node:
\draw[color=black] (0.000000,0.000000) -- (127.000000,0.000000);

% Line 9: c X b:owire
\begin{scope}[xshift=-44]
	\draw (77.500000,10.000000) -- (77.500000,0.000000);
	\draw (78.500000,10.000000) -- (78.500000,0.000000);

	\draw[fill=white] (78.000000, -0.000000) +(-45.000000:8.485281pt and 8.485281pt) -- +(45.000000:8.485281pt and 8.485281pt) -- +(135.000000:8.485281pt and 8.485281pt) -- +(225.000000:8.485281pt and 8.485281pt) -- cycle;
	\clip (78.000000, -0.000000) +(-45.000000:8.485281pt and 8.485281pt) -- +(45.000000:8.485281pt and 8.485281pt) -- +(135.000000:8.485281pt and 8.485281pt) -- +(225.000000:8.485281pt and 8.485281pt) -- cycle;
	\draw (78.000000, -0.000000) node {$X$};
\end{scope}
\node (m1) at (34.000000, 15.000000) {${\scriptstyle m_f}$};
\draw[->] (34,35)--(m1);
% Line 10: c Z b:owire # this is wrong (used to simplify manual editing of tikz)

% measurement lines
\draw (101.500000,10.000000) -- (101.500000,0.000000);
\draw (102.500000,10.000000) -- (102.500000,0.000000);
\begin{scope}
	\draw[fill=white] (102.000000, -0.000000) +(-45.000000:8.485281pt and 8.485281pt) -- +(45.000000:8.485281pt and 8.485281pt) -- +(135.000000:8.485281pt and 8.485281pt) -- +(225.000000:8.485281pt and 8.485281pt) -- cycle;
	\clip (102.000000, -0.000000) +(-45.000000:8.485281pt and 8.485281pt) -- +(45.000000:8.485281pt and 8.485281pt) -- +(135.000000:8.485281pt and 8.485281pt) -- +(225.000000:8.485281pt and 8.485281pt) -- cycle;
	\draw (102.000000, -0.000000) node {$Z$};
\end{scope}
\node (m2) at  (102.000000, 15.000000)  {${\scriptstyle m_u}$};
\draw[->] (102,50)--(m2);
\node at (105,37) {\rotatebox{-90}{\tiny send}};
\node at (37,29) {\rotatebox{-90}{\tiny send}};
% Done with gates; drawing ending labels
\draw[color=black] (124.000000,0.000000) node[right] {$|\psi\rangle$};
% Done with ending labels; drawing cut lines and comments
% Done with comments
\end{scope}
\end{tikzpicture}
	\caption{Quantum circuits illustrating fanout/unfanout and teleportation of a qubit in state $\ket\psi$ from node 1 to node 2 using an EPR pair. Quantum teleportation can be seen as a fanout to node 2, followed by an unfanout from node 2 to node 1. For teleportation, 1 EPR pair is used and 2 bits of classical information are sent to node 2.}
	\label{fig:teleport}
\end{figure}

When two nodes share a EPR pair, another qubit may be moved from one node to the other using the EPR pair and classical communication. The basic idea is to first fan out the qubit to the other node and to then remove (using measurement) the qubit from the first node. Fanout may be achieved using a parity measurement between the local EPR pair qubit and the qubit to send, followed by a conditional fix-up operation, as shown in \cref{fig:teleport}(a). We compute the parity using a CNOT gate between the qubit to send and the local share of the EPR pair, and then we measure the parity. If the outcome is 0, no further action is required. However, if the parity is 1, the other node must fix its "fanout" qubit (its share of the EPR pair) by flipping the qubit using an X gate. 

At this point, the qubit has been fanned out successfully to the second node. If both qubits were located on the same node, we could use a CNOT to reset the first qubit to $\ket0$, resulting in the qubit having moved from the original position to where the second qubit of the EPR pair was located. It turns out that the same is true in the remote setting: Because the CNOT resets the qubit to $\ket0$, we may use the principle of deferred measurement~\cite{nielsen2002quantum} to implement this uncomputing CNOT using just local gates and classical communication, as illustrated in \cref{fig:mcnot} and \cref{fig:teleport}(b): All that is needed is a measurement in the X-basis (apply Hadamard, then measure) and, if the outcome is 1, we apply a Z gate to the qubit on the second node. This completes the Unfanout(2$\rightarrow$1) section of the quantum circuit for teleportation in \cref{fig:teleport}(c). Note that only classical communication and no quantum communication is needed for the unfanout, see \cref{fig:teleport}(b).

\section{Distributed Quantum Computing}\label{sec:distrqc}

In order to run practical quantum applications, a fault tolerant quantum computer is necessary as current error rates for physical two-qubit gates are on the order of $10^{-3} - 10^{-4}$ \cite{lekitsch2017blueprint}, while practical applications in chemistry and cryptography require around $10^{10}$ Toffoli (or doubly-controlled NOT) gates~\cite{lee2020even,von2020quantum,haner2020improved,gidney2019factor}.
 
\textbf{Overhead from Fault Tolerance.} In order to store quantum information for the duration of computation without any errors, quantum error correction (QEC) uses many physical qubits to encode each logical qubit such that the error rates are low enough. Additionally, we require to implement quantum gates in a fault tolerant way to execute the quantum program on these logical qubits. Logical qubits and fault tolerant gates require a significant overhead in terms of physical qubits and runtime.
A modular quantum computer design might be beneficial to handle the large number of qubits, the cooling requirements for some technologies, and the necessary control electronics and hence we are considering distributed quantum computing in this work.
The logical clock cycle is optimistically assumed to be $10\mu s$ for midterm quantum computers in the paper by \citet{von2020quantum}. This number heavily depends on the choice of qubit technology and the physical error rates. For example state-of-the-art ion trap physical two-qubit gates take already $1.6 \mu s$ \cite{schafer2018fast} so the logical gate times for this technology will be slower than the estimated $10 \mu s$. \citet{lekitsch2017blueprint} estimate a logical gate time of 235$\mu s$ for ion traps which results in a runtime of 110 days to factor a 2048-bit number. Such slow logical gate times at least initially allow us to hide the latency of classical communication in a distributed setting.

A standard set of universal quantum fault-tolerant gates are the single-qubit Pauli, Hadamard, and $S$ gates, the single-qubit $T:=\sqrt{S}$ gate, and the two-qubit CNOT gate. Only the CNOT gates will require communication by either teleporting the involved qubits onto the same quantum node or by fanout of the control qubit to the other node. Both of which can be achieved by sharing logical EPR pairs. When using the surface code, which is currently viewed to be the most promising approach to enable fault-tolerant quantum computing, the most costly (local) operation is the $T$ gate. In contrast to other gates, $T$ gates require \textit{distillation}, which is performed in so-called \textit{magic state factories} \cite{bravyi2005universal}. The overhead due to these factories limits the parallelism on a fixed-size chip, since such factories are expected to require tens of thousands  of physical qubits when assuming physical error rates of $10^{-3}$~\cite{litinski2019magic}.

\textbf{Quantum-Coherent Interconnects.} The physical implementation of connecting different quantum nodes by, e.g., creating a distributed EPR pair, depends on the underlying technology. Using optical photons is a natural choice given their property to travel long distance with little perturbation. There is a large number of theoretical proposals \cite{PhysRevB.96.165312,xiang2017intracity, PhysRevA.89.022317} and also first experimental prototypes: optical photons have been used to demonstrate entanglement sharing between ion traps 20m apart from each other \cite{Hofmann72} or between atomic qubits \cite{moehring2007entanglement}. In superconducting transmon qubit architectures, it is necessary to convert microwave photons, which are are used to perform two-qubit operations locally, to optical photons \cite{forsch2020microwave}. However, such transducers are still challenging to build. Therefore, alternative approaches are also being pursued, e.g., directly coupling two quantum nodes with microwave photons in a cryogenic waveguide~\cite{magnard2020microwave, zhong2021deterministic}.

In addition to the physical implementation for entanglement sharing between nodes, a protocol for fault tolerance is required~\cite{PhysRevLett.78.4293, PhysRevLett.76.722, van2007communication, debone2020protocols}.

\textbf{Inter-Node Communication.} With entanglement sharing in place, it is possible to establish EPR pairs between nodes through the quantum-coherent interconnect. In turn, this enables quantum teleportation between nodes, thus allowing for sending and receiving quantum information with \textit{move} semantics.

However, we note that an additional mode of operation is possible: Instead of fully moving a qubit from one node to the other, the qubit may also be fanned out to the other node, thus exposing its value on multiple nodes at once. This is also referred to as an \textit{entangled copy}, which may be used, e.g., to reduce the delay of certain quantum circuits, see \cref{fig:fanout_simple} for an example.

In this second mode of operation, one may support all functionality of classical MPI. However, due to reversibility constraints, the inverse of each function must be available as well~\cite{bennett1973logical}. For example, reductions must be performed in a reversible manner. To this end, depending on the reduction operation, additional work qubits may be required. These must be stored and managed by the implementation until the inverse of the reduction is applied, allowing to uncompute these work qubits.

\section{Quantum MPI}

To allow programmers to express distributed algorithms in their quantum programming language of choice, we propose Quantum MPI (QMPI) -- a quantum extension of the classical message-passing interface (MPI) standard.

\subsection{Communicators and Interaction with MPI}

QMPI leverages MPI for classical communication. As such, the communication of classical and quantum data is completely separated: The first is handled by MPI, whereas QMPI handles the latter.

While some nodes in \texttt{MPI\_COMM\_WORLD} may be purely classical, such nodes must not be part of any communicator that is passed to a QMPI function. \texttt{QMPI\_COMM\_WORLD}, which is of type \texttt{MPI\_Comm}, contains all quantum (i.e., not purely classical) nodes. All quantum nodes must support classical MPI since otherwise, teleportation would not be possible, as it requires communicating classical bits.

\subsection{Datatypes}

Qubits may be allocated using \qmpi{Alloc\_qmem(n)}, where $n$ denotes the number of qubits to allocate. \qmpi{Alloc\_qmem} returns a \qmpi{QUBIT\_PTR} \texttt{qptr}, which points to the first qubit. Qubits may be deallocated using \qmpi{Free\_qmem}.

QMPI defines one basic quantum-specific datatype, \qmpi{QUBIT}, which represents a single quantum bit. Given that qubits will be a scarce resource initially, we leave the construction of more complex data types such as quantum integers and quantum floating-point numbers to the programmer: Such data types may be constructed from \qmpi{QUBIT} using \qmpi{Type\_*} functions such as \qmpi{Type\_contiguous}, as in classical MPI.

As we do not expect classical communication to be a bottleneck in the near term and in order to keep classical communication separate from quantum communication (the first using MPI, the second using QMPI), we do not allow for mixing of quantum and classical datatypes in the first version of QMPI. However, as protocols for quantum error correction and entanglement sharing are optimized, a tighter integration of QMPI with MPI may become critical to performance, and this restriction could thus be dropped if needed in a future version.

\subsection{EPR pairs}
The basic building block and most time consuming part for all quantum communication is the creation of EPR  pairs between qubits on the sending and receiving ranks. Established EPR pairs allow for higher-level communication primitives such as entangled copying (fanout) or moving (teleportation) of qubits between two nodes that share an EPR pair.

In order to request that an EPR pair be created between two nodes, each node invokes
\[
\qmpi{Prepare\_EPR(qubit, dest, tag, comm)},
\]
where \qmpiarg{qubit} is a fresh qubit in $\ket0$, \qmpiarg{dest} refers to the rank of a QMPI process running on the other node, \qmpiarg{tag} is the message tag, and \qmpiarg{comm} is the communicator (e.g., \qmpi{COMM\_WORLD}). Upon completion, the quantum state of the two qubits (located on different nodes) is $\frac{1}{\sqrt 2}(\ket{00}+\ket{11})$.

As is the case for other communication primitives, asynchronous versions (e.g., \qmpi{Iprepare\_EPR}) are available to allow for requesting EPR pairs ahead of time.

\subsection{General Point to Point Communication}

\renewcommand{\arraystretch}{1.3}
\begin{table*}
	\caption{Classical and quantum resources required for entangled copy, move, reduce, scan, and their respective inverse operations (or uncomputation) in brackets. Stated are the resources required per qubit in the message and for $N$ nodes (reduce/scan).}
	\label{tab:resources}
	\centering
	{\small
	\begin{tabular}{m{1.4cm}cccc}
		\toprule
		& copy [uncopy] & move [unmove] &  reduce [unreduce] & scan [unscan] \\
		\midrule 
		Quantum comm. (EPR pairs) & 1 [0] & 1 [1] & $N-1$ [0] & $N-1$ [0]\\
		Classical comm. (bits) & 1 [1] & 2 [2] & $N-1$ [$N-1$] & $N-1$ [$N-1$]\\
		\bottomrule
	\end{tabular}
	}
\end{table*}

\begin{table}
	\caption{Point to point communication primitives in QMPI. In addition to blocking, also non-blocking variants are available, as in classical MPI. Resource requirements are given in terms of entangled copy and move from \cref{tab:resources}. (a): Same as Sendrecv with move semantics, (b): Resources may already have been used.}
	\label{tab:p2p}
	\centering
	{\small
	\begin{tabular}{m{3.0cm}m{3.0cm}l}
		\toprule
		Operation & Reverse operation & Resources  \\
		\midrule 
		\raggedright QMPI\_Send, QMPI\_Bsend, QMPI\_Ssend, QMPI\_Rsend  & \raggedright QMPI\_Unsend, QMPI\_Bunsend QMPI\_Sunsend, QMPI\_Runsend & copy  \\
		
		QMPI\_Recv, QMPI\_Mrecv & QMPI\_Unrecv, QMPI\_Munrecv  & copy  \\
		
		QMPI\_Sendrecv & QMPI\_Unsendrecv & copy \\
		
		QMPI\_Sendrecv\_replace$^{(a)}$ &  QMPI\_Unsendrecv\_replace &  move  \\
		
		QMPI\_Cancel$^{(b)}$ & ---  \\
		\midrule
		QMPI\_Send\_move, QMPI\_Bsend\_move, QMPI\_Ssend\_move, QMPI\_Rsend\_move& QMPI\_Unsend\_move, QMPI\_Bunsend\_move, QMPI\_Sunsend\_move, QMPI\_Runsend\_move & move   \\
		
		QMPI\_Recv\_move, QMPI\_Mrecv\_move & QMPI\_Unrecv\_move, QMPI\_Munrecv\_move & move  \\
		\bottomrule
	\end{tabular}}
\end{table}

As discussed in \cref{sec:distrqc}, QMPI supports communication in two modes, one with copy semantics, the other with move semantics. Both modes rely on EPR pairs to move and to fan out qubits to other nodes. 
Qubits are moved from one node to another using quantum teleportation, whereas fanout exposes their values on multiple nodes simultaneously. QMPI provides functionality for fanning out and sending/receiving qubit (the latter with move semantics) via the two pairs of functions \qmpi{Send / Recv} and \qmpi{Send\_move / Recv\_move}. In addition, there are the inverses of \qmpi{Send / Recv}, denoted by \qmpi{Unsend / Unrecv}, respectively. The reason for this addition is that the uncomputation can be performed more efficiently: The qubit can simply be measured after applying a Hadamard gate and, if the outcome is nonzero, the other node must apply a Pauli Z gate to its qubit, as shown in \cref{fig:mcnot}(b). Therefore, uncomputing a fanned-out qubit can be achieved by communicating only a single bit of classical information without needing an EPR pair. The resource requirements for entangled copy/fanout, move, and their respective inverses can also be found in \cref{tab:resources}.
\cref{tab:p2p} lists all point-to-point primitives and the required resources in terms of the costs for entangled copy and move from \cref{tab:resources}.

\subsection{Collective Operations}

\begin{table}
	\caption{Collective communication in QMPI. In addition to the blocking calls, also non-blocking variants~\cite{standard-nbc} are available, as in classical MPI. Resource requirements are given in terms of entangled copy, move, reduce, and scan from \cref{tab:resources}. (a): For in-place: Move resources, (b): Operation must be reversible.}
	\label{tab:collectives}
	\centering
	{\small
	\begin{tabular}{m{2.7cm}m{3cm}l}
		\toprule
		Operation & Reverse operation & Resources  \\
		\midrule
		
		QMPI\_Bcast & QMPI\_Unbcast & copy \\
		
		QMPI\_Gather, QMPI\_Gatherv & QMPI\_Ungather, QMPI\_Ungatherv & copy \\
		
		QMPI\_Scatter, QMPI\_Scatterv & QMPI\_Unscatter, QMPI\_Unscatterv & copy \\
		
		QMPI\_Allgather, QMPI\_Allgatherv & QMPI\_Unallgather, QMPI\_Unallgatherv & copy \\
		
		QMPI\_Alltoall, QMPI\_Alltoallv, QMPI\_Alltoallw &QMPI\_Unalltoall, QMPI\_Unalltoallv, QMPI\_Unalltoallw & copy/move$^{(a)}$    \\
		QMPI\_Reduce & QMPI\_Unreduce & reduce$^{(b)}$  \\
		QMPI\_Allreduce & QMPI\_Unallreduce & reduce$^{(b)}$ + copy \\
		
		QMPI\_Reduce\_scatter, QMPI\_Reduce\_scatter- \_block & QMPI\_Unreduce\_scatter, QMPI\_Unreduce\_scatter- \_block & reduce$^{(b)}$\\
		
		QMPI\_Scan, QMPI\_Exscan & QMPI\_Unscan, QMPI\_Unexscan & scan$^{(b)}$ \\
		\midrule
		QMPI\_Gather\_move, QMPI\_Gatherv\_move &QMPI\_Ungather\_move, QMPI\_Ungatherv\_move &  move  \\
		
		QMPI\_Scatter\_move, QMPI\_Scatterv\_move &QMPI\_Unscatter\_move, QMPI\_Unscatterv\_move & move \\
		
		QMPI\_Alltoall\_move, QMPI\_Alltoallv\_move, QMPI\_Alltoallw\_move& QMPI\_Unalltoall\_move, QMPI\_Unalltoallv\_move, QMPI\_Unalltoallw\_move &  move  \\
		
		\bottomrule
	\end{tabular}}
\end{table}

In addition to general point-to-point communication, QMPI provides collective operations. \qmpi{Bcast} is an example of a simple QMPI collective implementing fanout. Its main purpose is to expose the value of a qubit on multiple nodes (and then uncomputing that value again with its inverse, \qmpi{Unbcast}), similar to copying a classical value with the corresponding MPI routine. In the quantum case,  collective communications allow even more optimizations beyond what is possible classically. In particular  \qmpi{Bcast} can be implemented with {\em constant} quantum time. As discussed by \citet[Theorem 17]{10.1145/3313276.3316404}, this can be done by first creating EPR pairs on all edges of a spanning tree of the nodes in the communicator as the only quantum communication step, which can be done in parallel in constant time. This is followed by local parity measurements among the entangled qubits at each node, the time for which is logarithmic in the maximum degree of a node in the spanning tree, which is typically a small constant. The last step consists of collective classical communication and computation to identify which qubits need to be changed by a Pauli $X$ gate. The logarithmic complexity of \qmpi{Bcast} is thus due to (fast) classical communication, while the (slow) quantum communication is of constant time.

\sloppy
Another collective operation, \qmpi{Scatter\_move} / \qmpi{Gather\_move} is an example of a QMPI collective with move semantics. A typical use case for this function is a section in the quantum algorithm where multiple rotation gates are applied to distinct qubits, all of which are located on the same node. In order to increase the number of local rotation factories per rotation qubit, the rotation qubits may be scatter-moved to separate nodes. After all rotations have been applied in parallel, the qubits may be gathered on the original node, allowing the computation to advance.

A QMPI collective with entangled copy semantics that is also worth a quick discussion is \qmpi{Reduce} (and its inverse \qmpi{Un\-reduce}). It differs from a classical MPI reduction only in that the reduction operation is reversible and that it must be uncomputed eventually (to free scratch space and to allow for interference in the quantum algorithm). In this first version, the QMPI implementation leaves all memory management to the user and \qmpi{Reduce} only accepts reversible operations\footnote{Future versions may support automatic compilation from a non-reversible implementation.}.

An example operation is \qmpi{PARITY}, which computes the parity of all qubits in the reduction. We note that there is a host of different methods that the QMPI implementation may choose from, depending on the situation (available scratch space, size of reduction, etc.). We refer to \cref{sec:examples} for a selection of different algorithms for computing the parity.

\cref{tab:collectives} shows a complete list of all QMPI collectives and the required resources in terms of entangled copy, move, reduce, and scan from \cref{tab:resources}.

\subsection{Communication Resources}

The tables with all point-to-point and collective operations give the resource requirements in terms of four basic primitives (and their inverses for uncomputing communicated data): entangled copy, move, reduce, and scan. \cref{tab:resources} can be used to translate from these basic primitives to the number of EPR pairs to be established, and the number of classical bits to be communicated. We note that the stated numbers for reduce and scan are representative of one particular implementation, and there are a host of different tradeoffs to consider in practice.

In particular, the stated numbers for reduce and scan are valid if sufficient logical qubits are available to store intermediate results. Using a linear communication schedule, both reduce and scan can be performed using a single output register per node and a total of $N-1$ EPR pairs per qubit to send, and uncomputation only requires classical communication. In contrast, a binary-tree reduction either requires more local storage, or intermediate results must be uncomputed, and later recomputed during \qmpi{Unreduce}, which also increases EPR pair usage. Similar considerations apply for the scan primitive.

For a more detailed discussion of the tradeoffs involved in optimizing collectives such as \qmpi{Bcast} and \qmpi{Reduce}, we refer the reader to \cref{sec:optimizing-bcast}.

\subsection{Future Extension: Persistent Requests}

Persistent communication requests allow further optimization beyond what is possible classically.  All required EPR pairs can be prepared before starting communication and, in particular, before the data to be sent is available. Point-to-point or collective quantum communication can then be performed with purely classical communication. This allows for overlaying quantum communication with computation performed {\it prior} to the communication start, which once more is impossible classically. Of course, this optimization is possible only if sufficient qubits are available to store the established EPR pairs and if there is sufficient time to establish all EPR pairs before the communication is started.

\section{The SENDQ Model}

Analogously to classical performance models such as the LogP model~\cite{culler1993logp}, whose parameters characterize the performance of the network interconnecting classical nodes, our SENDQ model captures the features of a distributed quantum computer that are most essential to performance.
We envision an architecture where multiple nodes are interconnected with both a classical and a quantum-coherent network, the latter of which may be used to send and receive quantum information. In particular, the quantum-coherent network is used to establish EPR pairs between two nodes.

We anticipate a relatively low logical clock speed for quantum computers due to the overhead introduced by the quantum error correction (QEC) protocol (cf.~Section~\ref{sec:distrqc}). As a consequence, we do not expect that classical communication will have a significant effect on performance and we thus choose to ignore classical communication in our model.

To account for optimizations that overlay communication with local computation, it is crucial to model the performance of both local and nonlocal operations. Our proposed model thus consists of two sets of parameters -- one to model (coherent) communication, and the other to model local computation.

In order to model the communication performance, we choose the following parameters.

\begin{enumerate}
	\item[$\bullet$ $S$:]The number of qubits used to store EPR pairs (per node).
	\item[$\bullet$ $E$:] (Upper bound on) the time it takes for a node to establish an EPR pair with any other node. Any node can be involved in at most one EPR pair creation at any point. We assume latencies are negligible.
	\item[$\bullet$ $N$:] The number of nodes.
\end{enumerate}

The local computation can be modeled using an abstract quantum circuit model that only considers width and depth of the circuit. The parameters are thus
\begin{enumerate}
	\item[$\bullet$ $D$:] The delay incurred due to local computation
	\item[$\bullet$ $Q$:] The number of logical qubits available for computation (per node)
\end{enumerate}

\subsection{Discussion of parameters}

We now briefly discuss the parameters that make up our performance model for distributed quantum computing.

\textbf{Parameter \textit S.} Our model of quantum communication includes a parameter for local storage, namely the number of logical qubits $S$ dedicated to buffering of EPR pairs.
This is different from classical performance models such as the LogP model \cite{culler1993logp}, which does not contain such parameters. This parameter is important because performance models with unlimited local storage allow for a simple and unrealistic exploit. Namely, all required EPR pairs may be shared and stored locally ahead of time. As a consequence, all quantum communication could then be implemented in constant time (ignoring the delay of classical communication), see \cref{sec:optimizing-bcast} for an in-depth explanation for the example of \qmpi{Bcast}.

\textbf{Parameter \textit E.} $E$ specifies the upper bound on the time it takes to establish a logical EPR pair with any other node, assuming exclusive communication. A logical EPR pair may be established by sharing many physical EPR pairs, followed by a distillation protocol. As we ignore latency, $E^{-1}$ can be seen as the EPR pair injection bandwidth per node into the quantum network.

\textbf{Parameter \textit N.}
The number of quantum nodes in the distributed quantum computer is denoted by $N$.

\textbf{Parameters \textit D and \textit Q.} Our model also includes parameters to model local compute as an integral part because logical qubits for computation can also be used for storing EPR pairs when unused. In general, only the total number of qubits $Q + S$ is constant on each node. Depending on the algorithm, one may choose $Q$ and $S$ to be fixed to a constant value in order to simplify the model even further. The delay $D$ can be specified in more detail if desired. For example, a common choice for a fault tolerant quantum computer is to ignore the delays of all gates and measurements except for the most costly rotation gates (arbitrary rotations and $T$ gates), as discussed in \cref{sec:distrqc}. Note that we consider the number of logical qubits per node $Q$ equivalent to the number of compute elements, i.e., the number of qubits onto which operations can be applied in parallel. This is due to the fact that current schemes for fault tolerance require full parallelism on all qubits in order to just store information and this parallelism can be used to apply gates.

For the applications presented in this paper, we assume that all parameters are constant throughout the execution of a given quantum algorithm.

\section{Prototype Implementation of QMPI}
We have implemented a QMPI prototype in C++ using MPI and multi-threading leveraging the C++ standard library. Our prototype supports a variety of standard quantum gates and the point-to-point as well as collective functions described in the previous section. The current implementation only supports qubit types, and no higher-level datatypes that may be constructed from qubits.

At the core of the library is a full state simulator that allows users to test and debug their distributed quantum algorithms.
To ensure that the state vector faithfully represents the quantum state of the distributed quantum computer at any point throughout the computation, all ranks forward quantum operations to rank 0, which then applies the operation to the state vector. Qubit allocations, deallocations, and measurements are handled similarly. Rank 0 runs a separate thread that waits to receive gate operations to execute. Consequently, all ranks (including rank 0) may be used in a quantum computation.

The following example shows how to establish an EPR pair between two QMPI ranks. The simulation output is as expected: Both ranks observe the same value when measuring their share of the EPR pair.

\begin{lstlisting}[style=cppstyle,escapechar=\%]
#include "qmpi.hpp"
#include <iostream>

using namespace QMPI;

int main() {
	QMPI_Init(0, 0);
	auto qubit = QMPI_Alloc_qmem(1); // allocate 1 qubit
	int rank;
	QMPI_Comm_rank(QMPI_COMM_WORLD, &rank);
	int dest = rank == 0 ? 1 : 0;
	// prepare EPR pair between rank and dest
	QMPI_Prepare_EPR(qubit, dest, 0, QMPI_COMM_WORLD);
	// measure the local qubit
	bool res = Measure(qubit);
	std::cout << rank << ": " << res << std::endl;
	QMPI_Free_qmem(qubit, 1); // free 1 qubit
	QMPI_Finalize();
	return 0;
}
\end{lstlisting}

In the next section, we describe more examples and we present the corresponding implementation based on our QMPI prototype.

\section{Applications}\label{sec:examples}

In this section, we show how quantum algorithms for applications from physics and chemistry may be implemented in QMPI and how the SENDQ model can be used to inform algorithmic decisions.

\subsection{Optimizing Collectives}
Collective operations allow hardware vendors to optimize these operations by taking into account their hardware parameters. In this section, we describe how to optimize \qmpi{Bcast}. In \cref{sec:chemistry}, we show an example of how to optimize \qmpi{Reduce} for systems with either one qubit per node dedicated to storing EPR paris ($S=1$) or systems with $S\geq 2$.

\textbf{Optimizing QMPI\_Bcast.}
\label{sec:optimizing-bcast}
\begin{figure}[t]
	\centering
	\begin{tikzpicture}[scale=1.200000,x=1pt,y=1pt]
\filldraw[color=white] (0.000000, -7.500000) rectangle (105.000000, 82.500000);

\node at (-25,90) {\footnotesize node 1:};
\fill[color=green!10!white,rounded corners] (-10,100) rectangle (110.000000, 82);
\node at (-25,65) {\footnotesize node 2:};
\fill[color=blue!20!white,rounded corners]  (-10,78) rectangle (110.000000, 52);
\node at (-25,35) {\footnotesize node 3:};
\fill[color=red!10!white,rounded corners]  (-10,48) rectangle (110.000000, 22);
\node at (-25,10) {\footnotesize node 4:};
\fill[color=yellow!20!white,rounded corners]  (-10,18) rectangle (110.000000, 0);
% Drawing wires
% Line 1: a0 a1 a2 a3 a4 a5 W
\draw[color=black] (0.000000,90.000000) -- (105.000000,90.000000);
% Line 1: a0 a1 a2 a3 a4 a5 W
\draw[color=black] (0.000000,70.000000) -- (105.000000,70.000000);
\draw[color=black] (0.000000,60.000000) -- (105.000000,60.000000);
% Line 1: a0 a1 a2 a3 a4 a5 W
\draw[color=black] (0.000000,40.000000) -- (105.000000,40.000000);
\draw[color=black] (0.000000,30.000000) -- (105.000000,30.000000);
% Line 1: a0 a1 a2 a3 a4 a5 W
\draw[color=black] (0.000000,10.000000) -- (105.000000,10.000000);
% Done with wires; drawing gates
% Line 3: a0 +a1
\draw[decorate,decoration={snake}] (9.000000,90.000000) -- (9.000000,70.000000);
\filldraw (9.000000, 90.000000) circle(1.500000pt);
\filldraw (9.000000, 70.000000) circle(1.500000pt);
% Line 4: a2 +a3
\draw[decorate,decoration={snake}]  (9.000000,30.000000) -- (9.000000,10.000000);
\filldraw (9.000000, 10.000000) circle(1.500000pt);
\filldraw (9.000000, 30.000000) circle(1.500000pt);

% Line 6: a1 +a2
\draw[decorate,decoration={snake}]  (27.000000,60.000000) -- (27.000000,40.000000);
\filldraw (27.000000, 60.000000) circle(1.500000pt);
\filldraw (27.000000, 40.000000) circle(1.500000pt);

% Line 9: a0 a2 a1 a3 a4 a5 G \rotatebox{90}{Measure \& fixup}
%\draw[decorate,decoration={snake}]  (48.000000,75.000000) -- (48.000000,0.000000);
\begin{scope}[yshift=12.5]
\draw[fill=white] (48.000000, 37.500000) +(-45.000000:8.485281pt and 61.518290pt) -- +(45.000000:8.485281pt and 61.518290pt) -- +(135.000000:8.485281pt and 61.518290pt) -- +(225.000000:8.485281pt and 61.518290pt) -- cycle;
\clip (48.000000, 37.500000) +(-45.000000:8.485281pt and 61.518290pt) -- +(45.000000:8.485281pt and 61.518290pt) -- +(135.000000:8.485281pt and 61.518290pt) -- +(225.000000:8.485281pt and 61.518290pt) -- cycle;
\draw (48.000000, 37.500000) node {\rotatebox{90}{Measure \& fixup}};
\end{scope}
% Line 11: =
\draw[fill=white,color=white] (66.000000, -6.000000) rectangle (81.000000, 100.000000);
\draw (73.500000, 50.00000) node {$=$};
% Line 13: a0 a1 a3 a5
\draw[decorate,decoration={snake}] (96.000000,90.000000) -- (96.000000, 70);
\draw[decorate,decoration={snake}] (96.000000,70) -- (96.000000, 40);
\draw[decorate,decoration={snake}] (96.000000, 40) -- (96.000000, 10);
\filldraw (96.000000, 90.000000) circle(1.500000pt);
\filldraw (96.000000, 70.000000) circle(1.500000pt);
\filldraw (96.000000, 40.000000) circle(1.500000pt);
\filldraw (96.000000, 10.000000) circle(1.500000pt);

\draw[decorate,decoration={brace,amplitude=5pt,mirror}] (5,-1) -- (32,-1);
\node at (18,-10) {$2E$ time};
% Done with gates; drawing ending labels
% Done with ending labels; drawing cut lines and comments
% Done with comments
\end{tikzpicture}
	\caption{Quantum circuit for establishing a cat state on $n=4$ nodes in constant quantum depth and classical $\mathcal O(\log n)$ depth. A reduction of measurement outcomes is required for computing the fixup operation for each node (see text).}
	\label{fig:catstate}
\end{figure}
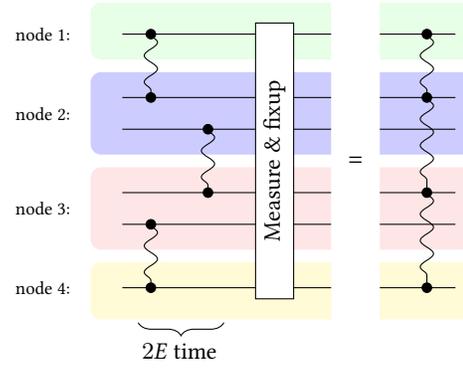
This first example presents a simple implementation of \qmpi{Bcast} in terms of \qmpi{Send / Recv} and shows how it can be analyzed and optimized using SENDQ. For simplicity, we assume that only one qubit is sent.

A log-depth implementation of broadcast can be achieved by constructing a binary tree of calls to \qmpi{Send / Recv}: In the $k$-th step (starting with $k=0$), $2^k$ nodes send the broadcast message to 1 other node, thus doubling the number of nodes that have received the message at every step. Since each node communicates with (at most) one node at every step, only one EPR pair must be established between each pair of nodes that communicates. As a result, $S=1$ is sufficient and the runtime of a broadcast is $E\lceil\log_2 N\rceil$.

This implementation may be optimized by realizing that a cat state, which is an $n$-qubit generalization of an EPR pair, that is,
\[
	\ket{\text{cat}(n)}:=\frac 1{\sqrt2}(\ket{\underbrace{0\cdots 0}_n}+\ket{\underbrace{1\cdots 1}_n}),
\]
can be prepared in constant depth~\cite{hoyer2005quantum}. In QMPI, $\ket{\text{cat}(n)}$ can be prepared by first connecting all $n$ nodes with EPR pairs along the edges of a spanning tree, and then combining the individual EPR pairs using a parity-measurement of the different EPR pair qubits on each node (no parity-measurement is performed on leaf nodes), see \cref{fig:catstate} for a simplified diagram of this process. The measurement outcomes are used to compute whether or not a given node must apply a Pauli X correction. Specifically, each node $k$ applies $X^{r_1\oplus\cdots\oplus r_{k-1}}$ to the qubit that will be part of the cat state, where $r_i$ denotes the outcome of the (in-place) parity measurement on node $i$, and $\bigoplus_{i=1}^{k-1} r_i$ can be computed with a classical \texttt{MPI\_Exscan}.

This procedure can be extended to an implementation of \qmpi{B\-cast} by also performing a parity measurement between the qubit to broadcast and the one EPR pair qubit on the root node. This implementation runs in quantum time
\[
	2E+D_M+D_F,
\]
where $D_M$ and $D_F$ denotes the time it takes to perform a local two-qubit parity measurement and to apply an X gate (the fixup operation), respectively. The classical \texttt{QMPI\_Exscan}, which is used to determine whether or not a local X gate correction is needed, can be performed in $\mathcal{O}(\log N)$ classical communication steps~\cite{sanders2006parallel}.

\subsection{Transverse-field Ising model}
\label{sec:TFIM}

In this second example, we show how to simulate the time evolution of a transverse-field Ising model (TFIM) with $n$ spins, whose Hamiltonian is given by
\[
	H_\text{TFIM} = \sum_{\langle i,j\rangle} J_{ij}\sigma_z^{(i)}\sigma_z^{(j)} - \sum_{i=0}^{n-1} \Gamma_i\sigma_x^{(i)},
\]
where $\sigma_x^{(i)}, \sigma_z^{(i)}$ denotes a Pauli X and Z, respectively, acting on spin index $i<n$, $J_{ij}$ denotes the coupling constant, and $\Gamma_i$ is the (local) strength of the transverse field. The first sum runs over all connected spins $i,j$, which we denote by $\langle i,j\rangle$.

Time evolution under this Hamiltonian can be used as a building block to solve optimization problems leveraging the adiabatic theorem~\cite{born1928beweis}: One first maps the optimization problem to a classical Ising model (thus defining the connectivity and the parameters $J_{ij}$). Then, starting with $J_{ij}=0,\Gamma_i=1$ and in the ground state of the corresponding Hamiltonian (which is $\ket{+}^{\otimes n}$), one slowly changes the parameters to $\Gamma_i=0$ and $J_{ij}$ equal to the computed values, aiming to remain in the groundstate of all intermediate Hamiltonians. Upon success, a final measurement of all qubits yields the solution of the optimization problem.

In the following, we assume linear nearest-neighbor connectivity for the spins and $J_{ij}=J,\Gamma_i=\Gamma$ for simplicity.
The time evolution operator for the Hamiltonian $H$ is given by $U(t)=e^{-itH}$, where $t$ is the time to evolve and $i^2=-1$. One possible approach to implement a TFIM simulation on a quantum computer is to first map each spin to a qubit. $U(t)$ may then be implemented by first decomposing it using a Trotter-Suzuki expansion. For example, a first-order approximation is
\[
	U(t)\approx \left(e^{i\delta t H_1}e^{i\delta tH_2}\right)^{\frac t{\delta t}},
\]
for small $\delta t$ and $H_1:=-J\sum_{\langle i,j\rangle} \sigma_z^{(i)}\sigma_z^{(j)}$, $H_2:=\Gamma\sum_i\sigma_x^{(i)}$. The individual terms in $H_1$ commute, and so do the terms within $H_2$. Therefore,
\[
	e^{-i\delta tH_1} = \prod_{\langle i,j\rangle} e^{-i\delta t J \sigma_z^{(i)}\sigma_z^{(j)}},\text{ and }e^{-i\delta tH_2} = \prod_{i} e^{i\delta t \Gamma \sigma_x^{(i)}}.
\]
Each term in the first product can be implemented by computing the parity between spin $i$ and $j$ using a CNOT gate, followed by a rotation gate $R_z(\theta)=e^{-0.5i\theta\sigma_z}$ and another CNOT gate to uncompute the parity. The terms in the second product are just rotation gates $R_x(\theta)=e^{-0.5i\theta\sigma_x}$ acting on qubit $i$.

The complete code for the simulation can be found in the appendix, see~\cref{lst:tfimcode}. 
While the prototype implementation uses blocking send/receive calls, we note that one would use an asynchronous version in practice: The EPR pairs could be established while applying the local operations.

\textbf{Analysis with SENDQ.} Each Trotter step requires $N$ EPR pairs, where $N$ denotes the number of nodes, and each node prepares an EPR pair with the two nodes that contain adjacent spins. We assume that rotation gates cannot be executed in parallel due to the cost (in space) of $T$-state factories. Since the delay of each rotation gate $D_R$ is much larger than the logical gate time, we ignore the cost of CNOTs. As a result, the delay of one Trotter step is approximately
\[
	D_{\text{Trotter}} = 2\frac nN D_R = 2 Q D_R,
\]
assuming that $n$ is divisible by $N$.

To ensure that communication is not a bottleneck (assuming asynchronous send/receive implementations), the time spent on local gates should be at least as large as the time it takes to establish two EPR pairs. For $S\geq 2$, this means that
\[
	D_{\text{Trotter}} \geq 2E.
\]

In turn, this allows us to inform our choice of the number of nodes $N$ if sufficient space is available per node to temporarily store the two EPR pairs: $N$ should be chosen such that
\[
	E^{-1} n D_R \geq N.
\]
If, on the other hand, space per node is a limiting factor and $S=1$ while $Q\geq2$, then one may return to the $S\geq 2$ case by increasing the number of nodes to $N\geq \lceil\frac{n}{Q-1}\rceil$.

We now address the case where increasing the number of nodes is not an option. Specifically, we show that our model correctly predicts an overhead for $S=1$ compared to $S\geq 2$ even with an optimized communication schedule that allows for halting local computations at any point, e.g., during execution of a local rotation gate.
With $S=1$, a request for EPR pair creation can only be initiated once the buffer qubit has been cleared. As a result, there is an additional delay $D_R$ between EPR pair creation requests because the rotation must be applied before the remote qubit can be unreceived. The delay per Trotter step with an optimized schedule for initiating EPR pair creation requests is thus
\[
	\max\left(D_{\text{Trotter}}, 2E+2D_R\right),
\]
in contrast to the $S\geq 2$ case, where the delay per Trotter step is $\max\left(D_{\text{Trotter}}, 2E\right)$.

This TFIM example shows that SENDQ can be used to model various tradeoffs in the implementation of a distributed quantum algorithm. Crucially, smaller $S$ results in longer runtimes, even if the communication schedule is optimized.

\subsection{Chemistry} \label{sec:chemistry}
Simulation of molecules is currently one of the most promising applications to first show a quantum advantage for a practical problem compared to classical supercomputers. The goal is to determine the energy eigenstates of molecules described by a Hamiltonian $H$. This then allows, for example, to investigate and optimize chemical catalysis \cite{von2020quantum}.

For large molecules the best quantum algorithms to find ground state energies are based on phase estimation of a unitary operator which depends only on the Hamiltonian of the molecule $H$. For a given molecule, the full quantum circuit is known at circuit compilation time, i.e., there are no branches in the program depending to measurements during runtime which influence performance. Hence, one can use expensive quantum circuit optimization techniques to reduce the quantum resources and increase performance ahead of time. We will highlight a few optimization possibilities for a distributed quantum computer.

We consider the algorithm of expressing the Hamiltonian $H$ of a molecule of interest in second quantization, expressed in a basis of $n$ spin-orbitals. This algorithm requires at least $n$ data qubits which might be distributed onto different nodes. We perform phase estimation on the time evolution operator of the system, $e^{-itH}$, which we implement using a Trotter-Suzuki expansion from \cref{sec:TFIM}. The majority of the algorithm is only one primitive operation, namely, a time evolution operator of the form:
\begin{equation}
\label{eq:local_term}
	e^{-it Z_{i_1} Z_{i_2}\cdots Z_{i_k}},  \quad \; \{i_1, \dots, i_k\} \subseteq \{0, \dots, n-1\}, t \in \mathbb{R}.
\end{equation}
The qubit indices and parameter $t$ involved in each of these operators depend on the molecule and on the choices of how to represent its Hamiltonian. 

\begin{figure}
	\includegraphics{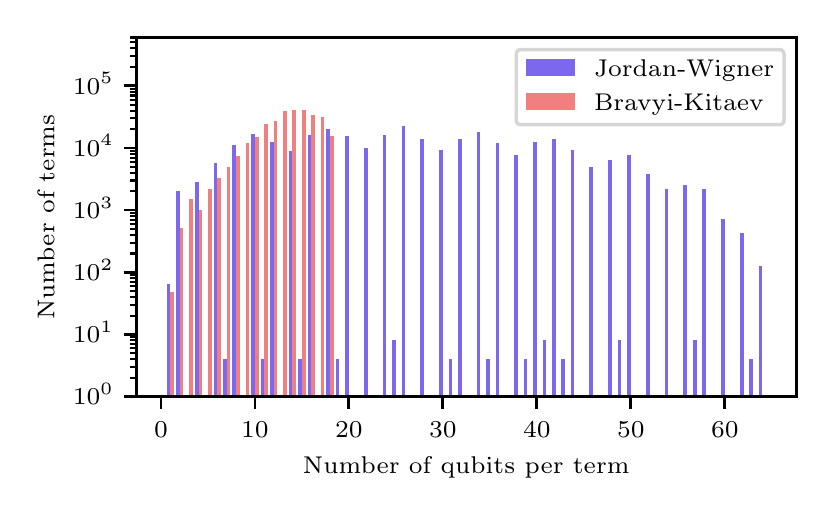}
	\caption{Mapping of the Hamiltonian representing a hydrogen ring with 32 atoms in the STO-3G basis set to a Hamiltonian acting on 64 qubits. The number of qubits involved in each term of the form defined by \cref{eq:local_term} is plotted as a histrogram for two different encoding methods.}
	\label{fig:mappings}
\end{figure}

\textbf{Analysis with SENDQ.} For a given molecule to be simulated, there are several choices to be made when mapping the problem to a quantum computer. In particular, different choices lead to different Hamiltonians, even if they all describe the same molecule. For example, the fermionic Hamiltonian needs to be transformed into a Hamiltonian that acts on qubits. This can be achieved using the Jordan-Wigner transformation \cite{Jordan1928,ortiz2001quantum,somma2002simulating}, the Bravyi-Kitaev encoding \cite{BRAVYI2002210}, or by using more than $n$ data qubits \cite{PhysRevA.94.030301}. For example, a Jordan-Wigner transformation will result in operations as in \cref{eq:local_term}, which may act on all data qubits. In contrast, the operators resulting from a Bravyi-Kitaev encoding only act on at most $\mathcal{O}(\log n)$ qubits, which may lead to savings in a distributed setting, at least without considering further optimizations to the Jordan-Wigner approach. The mapping differences are illustrated in \cref{fig:mappings} for the example of a hydrogen ring (data was generated using Refs.~\cite{PYSCF, steiger2018projectq, mcclean2020}).

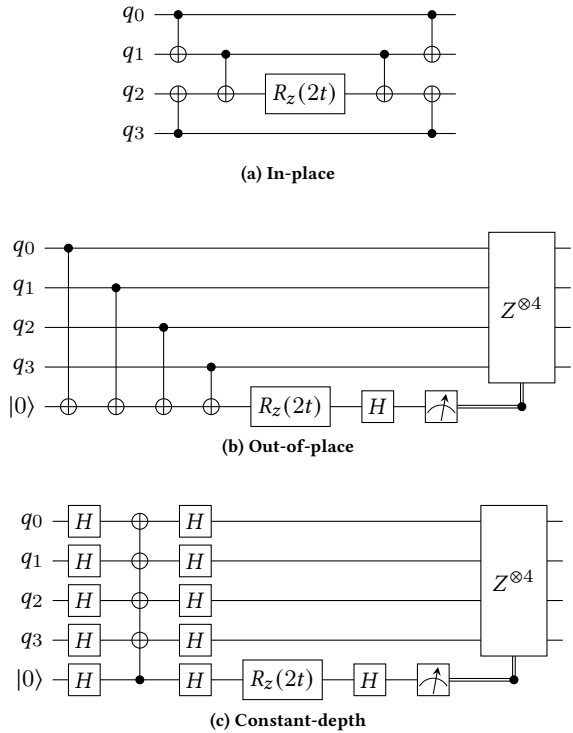
\begin{figure}
	\centering
	\subfloat[In-place]{\begin{tikzpicture}[scale=1.000000,x=1pt,y=1pt]
\filldraw[color=white] (0.000000, -7.500000) rectangle (114.000000, 52.500000);
% Drawing wires
% Line 2: q0 W q_0
\draw[color=black] (0.000000,45.000000) -- (114.000000,45.000000);
\draw[color=black] (0.000000,45.000000) node[left] {$q_0$};
% Line 3: q1 W q_1
\draw[color=black] (0.000000,30.000000) -- (114.000000,30.000000);
\draw[color=black] (0.000000,30.000000) node[left] {$q_1$};
% Line 4: q2 W q_2
\draw[color=black] (0.000000,15.000000) -- (114.000000,15.000000);
\draw[color=black] (0.000000,15.000000) node[left] {$q_2$};
% Line 5: q3 W q_3
\draw[color=black] (0.000000,0.000000) -- (114.000000,0.000000);
\draw[color=black] (0.000000,0.000000) node[left] {$q_3$};
% Done with wires; drawing gates
% Line 7: +q1 q0
\draw (9.000000,45.000000) -- (9.000000,30.000000);
\begin{scope}
\draw[fill=white] (9.000000, 30.000000) circle(3.000000pt);
\clip (9.000000, 30.000000) circle(3.000000pt);
\draw (6.000000, 30.000000) -- (12.000000, 30.000000);
\draw (9.000000, 27.000000) -- (9.000000, 33.000000);
\end{scope}
\filldraw (9.000000, 45.000000) circle(1.500000pt);
% Line 8: +q2 q3
\draw (9.000000,15.000000) -- (9.000000,0.000000);
\begin{scope}
\draw[fill=white] (9.000000, 15.000000) circle(3.000000pt);
\clip (9.000000, 15.000000) circle(3.000000pt);
\draw (6.000000, 15.000000) -- (12.000000, 15.000000);
\draw (9.000000, 12.000000) -- (9.000000, 18.000000);
\end{scope}
\filldraw (9.000000, 0.000000) circle(1.500000pt);
% Line 9: +q2 q1
\draw (27.000000,30.000000) -- (27.000000,15.000000);
\begin{scope}
\draw[fill=white] (27.000000, 15.000000) circle(3.000000pt);
\clip (27.000000, 15.000000) circle(3.000000pt);
\draw (24.000000, 15.000000) -- (30.000000, 15.000000);
\draw (27.000000, 12.000000) -- (27.000000, 18.000000);
\end{scope}
\filldraw (27.000000, 30.000000) circle(1.500000pt);
% Line 11: q2 G $R_z(2t)$ width=30 height=15
\begin{scope}
\draw[fill=white] (57.000000, 15.000000) +(-45.000000:21.213203pt and 10.606602pt) -- +(45.000000:21.213203pt and 10.606602pt) -- +(135.000000:21.213203pt and 10.606602pt) -- +(225.000000:21.213203pt and 10.606602pt) -- cycle;
\clip (57.000000, 15.000000) +(-45.000000:21.213203pt and 10.606602pt) -- +(45.000000:21.213203pt and 10.606602pt) -- +(135.000000:21.213203pt and 10.606602pt) -- +(225.000000:21.213203pt and 10.606602pt) -- cycle;
\draw (57.000000, 15.000000) node {$R_z(2t)$};
\end{scope}
% Line 13: +q2 q1
\draw (87.000000,30.000000) -- (87.000000,15.000000);
\begin{scope}
\draw[fill=white] (87.000000, 15.000000) circle(3.000000pt);
\clip (87.000000, 15.000000) circle(3.000000pt);
\draw (84.000000, 15.000000) -- (90.000000, 15.000000);
\draw (87.000000, 12.000000) -- (87.000000, 18.000000);
\end{scope}
\filldraw (87.000000, 30.000000) circle(1.500000pt);
% Line 14: +q2 q3
\draw (105.000000,15.000000) -- (105.000000,0.000000);
\begin{scope}
\draw[fill=white] (105.000000, 15.000000) circle(3.000000pt);
\clip (105.000000, 15.000000) circle(3.000000pt);
\draw (102.000000, 15.000000) -- (108.000000, 15.000000);
\draw (105.000000, 12.000000) -- (105.000000, 18.000000);
\end{scope}
\filldraw (105.000000, 0.000000) circle(1.500000pt);
% Line 15: +q1 q0
\draw (105.000000,45.000000) -- (105.000000,30.000000);
\begin{scope}
\draw[fill=white] (105.000000, 30.000000) circle(3.000000pt);
\clip (105.000000, 30.000000) circle(3.000000pt);
\draw (102.000000, 30.000000) -- (108.000000, 30.000000);
\draw (105.000000, 27.000000) -- (105.000000, 33.000000);
\end{scope}
\filldraw (105.000000, 45.000000) circle(1.500000pt);
% Done with gates; drawing ending labels
% Done with ending labels; drawing cut lines and comments
% Done with comments
\end{tikzpicture}}\\ \vspace{5pt}
	\subfloat[Out-of-place]{\begin{tikzpicture}[scale=1.000000,x=1pt,y=1pt]
\filldraw[color=white] (0.000000, -7.500000) rectangle (199.000000, 67.500000);
% Drawing wires
% Line 2: q5 W q_0
\draw[color=black] (0.000000,60.000000) -- (199.000000,60.000000);
\draw[color=black] (0.000000,60.000000) node[left] {$q_0$};
% Line 3: q6 W q_1
\draw[color=black] (0.000000,45.000000) -- (199.000000,45.000000);
\draw[color=black] (0.000000,45.000000) node[left] {$q_1$};
% Line 4: q7 W q_2
\draw[color=black] (0.000000,30.000000) -- (199.000000,30.000000);
\draw[color=black] (0.000000,30.000000) node[left] {$q_2$};
% Line 5: q8 W q_3
\draw[color=black] (0.000000,15.000000) -- (199.000000,15.000000);
\draw[color=black] (0.000000,15.000000) node[left] {$q_3$};
% Line 6: q9 W |0\rangle
\draw[color=black] (0.000000,0.000000) -- (150.000000,0.000000);
\draw[color=black] (150.000000,-0.500000) -- (180.500000,-0.500000);
\draw[color=black] (150.000000,0.500000) -- (180.500000,0.500000);
\draw[color=black] (0.000000,0.000000) node[left] {$|0\rangle$};
% Done with wires; drawing gates
% Line 8: +q9 q5
\draw (9.000000,60.000000) -- (9.000000,0.000000);
\begin{scope}
\draw[fill=white] (9.000000, 0.000000) circle(3.000000pt);
\clip (9.000000, 0.000000) circle(3.000000pt);
\draw (6.000000, 0.000000) -- (12.000000, 0.000000);
\draw (9.000000, -3.000000) -- (9.000000, 3.000000);
\end{scope}
\filldraw (9.000000, 60.000000) circle(1.500000pt);
% Line 9: +q9 q6
\draw (27.000000,45.000000) -- (27.000000,0.000000);
\begin{scope}
\draw[fill=white] (27.000000, 0.000000) circle(3.000000pt);
\clip (27.000000, 0.000000) circle(3.000000pt);
\draw (24.000000, 0.000000) -- (30.000000, 0.000000);
\draw (27.000000, -3.000000) -- (27.000000, 3.000000);
\end{scope}
\filldraw (27.000000, 45.000000) circle(1.500000pt);
% Line 10: +q9 q7
\draw (45.000000,30.000000) -- (45.000000,0.000000);
\begin{scope}
\draw[fill=white] (45.000000, 0.000000) circle(3.000000pt);
\clip (45.000000, 0.000000) circle(3.000000pt);
\draw (42.000000, 0.000000) -- (48.000000, 0.000000);
\draw (45.000000, -3.000000) -- (45.000000, 3.000000);
\end{scope}
\filldraw (45.000000, 30.000000) circle(1.500000pt);
% Line 11: +q9 q8
\draw (63.000000,15.000000) -- (63.000000,0.000000);
\begin{scope}
\draw[fill=white] (63.000000, 0.000000) circle(3.000000pt);
\clip (63.000000, 0.000000) circle(3.000000pt);
\draw (60.000000, 0.000000) -- (66.000000, 0.000000);
\draw (63.000000, -3.000000) -- (63.000000, 3.000000);
\end{scope}
\filldraw (63.000000, 15.000000) circle(1.500000pt);
% Line 13: q9 G $R_z(2t)$ width=30 height=15
\begin{scope}
\draw[fill=white] (93.000000, -0.000000) +(-45.000000:21.213203pt and 10.606602pt) -- +(45.000000:21.213203pt and 10.606602pt) -- +(135.000000:21.213203pt and 10.606602pt) -- +(225.000000:21.213203pt and 10.606602pt) -- cycle;
\clip (93.000000, -0.000000) +(-45.000000:21.213203pt and 10.606602pt) -- +(45.000000:21.213203pt and 10.606602pt) -- +(135.000000:21.213203pt and 10.606602pt) -- +(225.000000:21.213203pt and 10.606602pt) -- cycle;
\draw (93.000000, -0.000000) node {$R_z(2t)$};
\end{scope}
% Line 15: q9 H
\begin{scope}
\draw[fill=white] (126.000000, -0.000000) +(-45.000000:8.485281pt and 8.485281pt) -- +(45.000000:8.485281pt and 8.485281pt) -- +(135.000000:8.485281pt and 8.485281pt) -- +(225.000000:8.485281pt and 8.485281pt) -- cycle;
\clip (126.000000, -0.000000) +(-45.000000:8.485281pt and 8.485281pt) -- +(45.000000:8.485281pt and 8.485281pt) -- +(135.000000:8.485281pt and 8.485281pt) -- +(225.000000:8.485281pt and 8.485281pt) -- cycle;
\draw (126.000000, -0.000000) node {$H$};
\end{scope}
% Line 16: q9 M
\draw[fill=white] (144.000000, -6.000000) rectangle (156.000000, 6.000000);
\draw[very thin] (150.000000, 0.600000) arc (90:150:6.000000pt);
\draw[very thin] (150.000000, 0.600000) arc (90:30:6.000000pt);
\draw[->,>=stealth] (150.000000, -5.400000) -- +(80:10.392305pt);
% Line 18: q5 q6 q7 q8 G $Z^{\otimes 4}$ width=25 q9:owire
\draw (180.500000,60.000000) -- (180.500000,15.000000);
\draw (180.000000,15.000000) -- (180.000000,0.000000);
\draw (181.000000,15.000000) -- (181.000000,0.000000);
\begin{scope}
\draw[fill=white] (180.500000, 37.500000) +(-45.000000:17.677670pt and 40.305087pt) -- +(45.000000:17.677670pt and 40.305087pt) -- +(135.000000:17.677670pt and 40.305087pt) -- +(225.000000:17.677670pt and 40.305087pt) -- cycle;
\clip (180.500000, 37.500000) +(-45.000000:17.677670pt and 40.305087pt) -- +(45.000000:17.677670pt and 40.305087pt) -- +(135.000000:17.677670pt and 40.305087pt) -- +(225.000000:17.677670pt and 40.305087pt) -- cycle;
\draw (180.500000, 37.500000) node {$Z^{\otimes 4}$};
\end{scope}
\filldraw (180.500000, 0.000000) circle(1.500000pt);
% Done with gates; drawing ending labels
% Done with ending labels; drawing cut lines and comments
% Done with comments
\end{tikzpicture}}\\ \vspace{5pt}
	\subfloat[Constant-depth]{\begin{tikzpicture}[scale=1.000000,x=1pt,y=1pt]
\filldraw[color=white] (0.000000, -7.500000) rectangle (193.000000, 67.500000);
% Drawing wires
% Line 2: q12 W q_0
\draw[color=black] (0.000000,60.000000) -- (193.000000,60.000000);
\draw[color=black] (0.000000,60.000000) node[left] {$q_0$};
% Line 3: q13 W q_1
\draw[color=black] (0.000000,45.000000) -- (193.000000,45.000000);
\draw[color=black] (0.000000,45.000000) node[left] {$q_1$};
% Line 4: q14 W q_2
\draw[color=black] (0.000000,30.000000) -- (193.000000,30.000000);
\draw[color=black] (0.000000,30.000000) node[left] {$q_2$};
% Line 5: q15 W q_3
\draw[color=black] (0.000000,15.000000) -- (193.000000,15.000000);
\draw[color=black] (0.000000,15.000000) node[left] {$q_3$};
% Line 6: q16 W |0\rangle
\draw[color=black] (0.000000,0.000000) -- (144.000000,0.000000);
\draw[color=black] (144.000000,-0.500000) -- (174.500000,-0.500000);
\draw[color=black] (144.000000,0.500000) -- (174.500000,0.500000);
\draw[color=black] (0.000000,0.000000) node[left] {$|0\rangle$};
% Done with wires; drawing gates
% Line 8: q12 H
\begin{scope}
\draw[fill=white] (12.000000, 60.000000) +(-45.000000:8.485281pt and 8.485281pt) -- +(45.000000:8.485281pt and 8.485281pt) -- +(135.000000:8.485281pt and 8.485281pt) -- +(225.000000:8.485281pt and 8.485281pt) -- cycle;
\clip (12.000000, 60.000000) +(-45.000000:8.485281pt and 8.485281pt) -- +(45.000000:8.485281pt and 8.485281pt) -- +(135.000000:8.485281pt and 8.485281pt) -- +(225.000000:8.485281pt and 8.485281pt) -- cycle;
\draw (12.000000, 60.000000) node {$H$};
\end{scope}
% Line 9: q13 H
\begin{scope}
\draw[fill=white] (12.000000, 45.000000) +(-45.000000:8.485281pt and 8.485281pt) -- +(45.000000:8.485281pt and 8.485281pt) -- +(135.000000:8.485281pt and 8.485281pt) -- +(225.000000:8.485281pt and 8.485281pt) -- cycle;
\clip (12.000000, 45.000000) +(-45.000000:8.485281pt and 8.485281pt) -- +(45.000000:8.485281pt and 8.485281pt) -- +(135.000000:8.485281pt and 8.485281pt) -- +(225.000000:8.485281pt and 8.485281pt) -- cycle;
\draw (12.000000, 45.000000) node {$H$};
\end{scope}
% Line 10: q14 H
\begin{scope}
\draw[fill=white] (12.000000, 30.000000) +(-45.000000:8.485281pt and 8.485281pt) -- +(45.000000:8.485281pt and 8.485281pt) -- +(135.000000:8.485281pt and 8.485281pt) -- +(225.000000:8.485281pt and 8.485281pt) -- cycle;
\clip (12.000000, 30.000000) +(-45.000000:8.485281pt and 8.485281pt) -- +(45.000000:8.485281pt and 8.485281pt) -- +(135.000000:8.485281pt and 8.485281pt) -- +(225.000000:8.485281pt and 8.485281pt) -- cycle;
\draw (12.000000, 30.000000) node {$H$};
\end{scope}
% Line 11: q15 H
\begin{scope}
\draw[fill=white] (12.000000, 15.000000) +(-45.000000:8.485281pt and 8.485281pt) -- +(45.000000:8.485281pt and 8.485281pt) -- +(135.000000:8.485281pt and 8.485281pt) -- +(225.000000:8.485281pt and 8.485281pt) -- cycle;
\clip (12.000000, 15.000000) +(-45.000000:8.485281pt and 8.485281pt) -- +(45.000000:8.485281pt and 8.485281pt) -- +(135.000000:8.485281pt and 8.485281pt) -- +(225.000000:8.485281pt and 8.485281pt) -- cycle;
\draw (12.000000, 15.000000) node {$H$};
\end{scope}
% Line 12: q16 H
\begin{scope}
\draw[fill=white] (12.000000, -0.000000) +(-45.000000:8.485281pt and 8.485281pt) -- +(45.000000:8.485281pt and 8.485281pt) -- +(135.000000:8.485281pt and 8.485281pt) -- +(225.000000:8.485281pt and 8.485281pt) -- cycle;
\clip (12.000000, -0.000000) +(-45.000000:8.485281pt and 8.485281pt) -- +(45.000000:8.485281pt and 8.485281pt) -- +(135.000000:8.485281pt and 8.485281pt) -- +(225.000000:8.485281pt and 8.485281pt) -- cycle;
\draw (12.000000, -0.000000) node {$H$};
\end{scope}
% Line 14: +q12 +q13 +q14 +q15 q16
\draw (33.000000,60.000000) -- (33.000000,0.000000);
\begin{scope}
\draw[fill=white] (33.000000, 60.000000) circle(3.000000pt);
\clip (33.000000, 60.000000) circle(3.000000pt);
\draw (30.000000, 60.000000) -- (36.000000, 60.000000);
\draw (33.000000, 57.000000) -- (33.000000, 63.000000);
\end{scope}
\begin{scope}
\draw[fill=white] (33.000000, 45.000000) circle(3.000000pt);
\clip (33.000000, 45.000000) circle(3.000000pt);
\draw (30.000000, 45.000000) -- (36.000000, 45.000000);
\draw (33.000000, 42.000000) -- (33.000000, 48.000000);
\end{scope}
\begin{scope}
\draw[fill=white] (33.000000, 30.000000) circle(3.000000pt);
\clip (33.000000, 30.000000) circle(3.000000pt);
\draw (30.000000, 30.000000) -- (36.000000, 30.000000);
\draw (33.000000, 27.000000) -- (33.000000, 33.000000);
\end{scope}
\begin{scope}
\draw[fill=white] (33.000000, 15.000000) circle(3.000000pt);
\clip (33.000000, 15.000000) circle(3.000000pt);
\draw (30.000000, 15.000000) -- (36.000000, 15.000000);
\draw (33.000000, 12.000000) -- (33.000000, 18.000000);
\end{scope}
\filldraw (33.000000, 0.000000) circle(1.500000pt);
% Line 16: q12 H
\begin{scope}
\draw[fill=white] (54.000000, 60.000000) +(-45.000000:8.485281pt and 8.485281pt) -- +(45.000000:8.485281pt and 8.485281pt) -- +(135.000000:8.485281pt and 8.485281pt) -- +(225.000000:8.485281pt and 8.485281pt) -- cycle;
\clip (54.000000, 60.000000) +(-45.000000:8.485281pt and 8.485281pt) -- +(45.000000:8.485281pt and 8.485281pt) -- +(135.000000:8.485281pt and 8.485281pt) -- +(225.000000:8.485281pt and 8.485281pt) -- cycle;
\draw (54.000000, 60.000000) node {$H$};
\end{scope}
% Line 17: q13 H
\begin{scope}
\draw[fill=white] (54.000000, 45.000000) +(-45.000000:8.485281pt and 8.485281pt) -- +(45.000000:8.485281pt and 8.485281pt) -- +(135.000000:8.485281pt and 8.485281pt) -- +(225.000000:8.485281pt and 8.485281pt) -- cycle;
\clip (54.000000, 45.000000) +(-45.000000:8.485281pt and 8.485281pt) -- +(45.000000:8.485281pt and 8.485281pt) -- +(135.000000:8.485281pt and 8.485281pt) -- +(225.000000:8.485281pt and 8.485281pt) -- cycle;
\draw (54.000000, 45.000000) node {$H$};
\end{scope}
% Line 18: q14 H
\begin{scope}
\draw[fill=white] (54.000000, 30.000000) +(-45.000000:8.485281pt and 8.485281pt) -- +(45.000000:8.485281pt and 8.485281pt) -- +(135.000000:8.485281pt and 8.485281pt) -- +(225.000000:8.485281pt and 8.485281pt) -- cycle;
\clip (54.000000, 30.000000) +(-45.000000:8.485281pt and 8.485281pt) -- +(45.000000:8.485281pt and 8.485281pt) -- +(135.000000:8.485281pt and 8.485281pt) -- +(225.000000:8.485281pt and 8.485281pt) -- cycle;
\draw (54.000000, 30.000000) node {$H$};
\end{scope}
% Line 19: q15 H
\begin{scope}
\draw[fill=white] (54.000000, 15.000000) +(-45.000000:8.485281pt and 8.485281pt) -- +(45.000000:8.485281pt and 8.485281pt) -- +(135.000000:8.485281pt and 8.485281pt) -- +(225.000000:8.485281pt and 8.485281pt) -- cycle;
\clip (54.000000, 15.000000) +(-45.000000:8.485281pt and 8.485281pt) -- +(45.000000:8.485281pt and 8.485281pt) -- +(135.000000:8.485281pt and 8.485281pt) -- +(225.000000:8.485281pt and 8.485281pt) -- cycle;
\draw (54.000000, 15.000000) node {$H$};
\end{scope}
% Line 20: q16 H
\begin{scope}
\draw[fill=white] (54.000000, -0.000000) +(-45.000000:8.485281pt and 8.485281pt) -- +(45.000000:8.485281pt and 8.485281pt) -- +(135.000000:8.485281pt and 8.485281pt) -- +(225.000000:8.485281pt and 8.485281pt) -- cycle;
\clip (54.000000, -0.000000) +(-45.000000:8.485281pt and 8.485281pt) -- +(45.000000:8.485281pt and 8.485281pt) -- +(135.000000:8.485281pt and 8.485281pt) -- +(225.000000:8.485281pt and 8.485281pt) -- cycle;
\draw (54.000000, -0.000000) node {$H$};
\end{scope}
% Line 22: q16 G $R_z(2t)$ width=30 height=15
\begin{scope}
\draw[fill=white] (87.000000, -0.000000) +(-45.000000:21.213203pt and 10.606602pt) -- +(45.000000:21.213203pt and 10.606602pt) -- +(135.000000:21.213203pt and 10.606602pt) -- +(225.000000:21.213203pt and 10.606602pt) -- cycle;
\clip (87.000000, -0.000000) +(-45.000000:21.213203pt and 10.606602pt) -- +(45.000000:21.213203pt and 10.606602pt) -- +(135.000000:21.213203pt and 10.606602pt) -- +(225.000000:21.213203pt and 10.606602pt) -- cycle;
\draw (87.000000, -0.000000) node {$R_z(2t)$};
\end{scope}
% Line 24: q16 H
\begin{scope}
\draw[fill=white] (120.000000, -0.000000) +(-45.000000:8.485281pt and 8.485281pt) -- +(45.000000:8.485281pt and 8.485281pt) -- +(135.000000:8.485281pt and 8.485281pt) -- +(225.000000:8.485281pt and 8.485281pt) -- cycle;
\clip (120.000000, -0.000000) +(-45.000000:8.485281pt and 8.485281pt) -- +(45.000000:8.485281pt and 8.485281pt) -- +(135.000000:8.485281pt and 8.485281pt) -- +(225.000000:8.485281pt and 8.485281pt) -- cycle;
\draw (120.000000, -0.000000) node {$H$};
\end{scope}
% Line 25: q16 M
\draw[fill=white] (138.000000, -6.000000) rectangle (150.000000, 6.000000);
\draw[very thin] (144.000000, 0.600000) arc (90:150:6.000000pt);
\draw[very thin] (144.000000, 0.600000) arc (90:30:6.000000pt);
\draw[->,>=stealth] (144.000000, -5.400000) -- +(80:10.392305pt);
% Line 27: q12 q13 q14 q15 G $Z^{\otimes 4}$ width=25 q16:owire
\draw (174.500000,60.000000) -- (174.500000,15.000000);
\draw (174.000000,15.000000) -- (174.000000,0.000000);
\draw (175.000000,15.000000) -- (175.000000,0.000000);
\begin{scope}
\draw[fill=white] (174.500000, 37.500000) +(-45.000000:17.677670pt and 40.305087pt) -- +(45.000000:17.677670pt and 40.305087pt) -- +(135.000000:17.677670pt and 40.305087pt) -- +(225.000000:17.677670pt and 40.305087pt) -- cycle;
\clip (174.500000, 37.500000) +(-45.000000:17.677670pt and 40.305087pt) -- +(45.000000:17.677670pt and 40.305087pt) -- +(135.000000:17.677670pt and 40.305087pt) -- +(225.000000:17.677670pt and 40.305087pt) -- cycle;
\draw (174.500000, 37.500000) node {$Z^{\otimes 4}$};
\end{scope}
\filldraw (174.500000, 0.000000) circle(1.500000pt);
% Done with gates; drawing ending labels
% Done with ending labels; drawing cut lines and comments
% Done with comments
\end{tikzpicture}} \vspace{5pt}
	\caption{Three different methods to implement $e^{-itZ_0Z_1 \cdots Z_{k-1}}$ for $k=4$.}
	\label{fig:chem_circuits}
\end{figure}

Once the encoding has been fixed, the individual operators must be implemented in terms of quantum gates. Here, we discuss the tradeoffs of three different approaches to implementing operators of the form given by \cref{eq:local_term}. For simplicity, we assume that each of the qubits involved is on a different node and that rotation gates take much longer to execute than measurements and other (local) quantum gates, allowing us to ignore the latter. \cref{fig:chem_circuits} illustrates the three approaches for an operator acting on $k=4$ qubits. Each of these circuits consists of the same three subroutines: a parity computation of all involved qubits into a target qubit, a single qubit rotation $R_z(2t)$ on that target qubit, and a final uncomputation of the parity.

The circuit in \cref{fig:chem_circuits}(a) computes the parity in place using a binary tree of distributed CNOT gates. Consequently, the full circuit requires $2(k-1)$ EPR pairs and has a runtime of
\[
	2 E\lceil\log_2k\rceil + D_R,
\]
where $D_R$ is the delay to execute one rotation gate. 

The circuit in \cref{fig:chem_circuits}(b) computes the parity into an auxiliary qubit. The downside is that the distributed CNOT gates now must be performed serially (unless more auxiliary qubits are available), but the uncomputation can be performed using only classical communication (see \cref{fig:mcnot}). As a result, only $k$ EPR pairs are required, but the circuit delay is
\[
	Ek + D_R.
\]
The parity computation of both \cref{fig:chem_circuits}(a) and (b) can be expressed as a reduction in QMPI, i.e., with a call to \qmpi{Reduce}.

In contrast to the first two circuits, \cref{fig:chem_circuits}(c) illustrates that a constant-depth implementation is possible in quantum computing using a parallel implementation of the multi-target CNOT. Specifically, this involves fanning out the control qubit using \qmpi{Bcast} to each node, which requires $k$ EPR pairs to establish a cat state, see \cref{sec:optimizing-bcast}, and, thus, $S\geq2$ is needed~\cite{hoyer2005quantum,kliuchnikov2021}. The delay of this constant-depth implementation is
\[
	2E + D_R.
\]

\begin{figure}
	\includegraphics{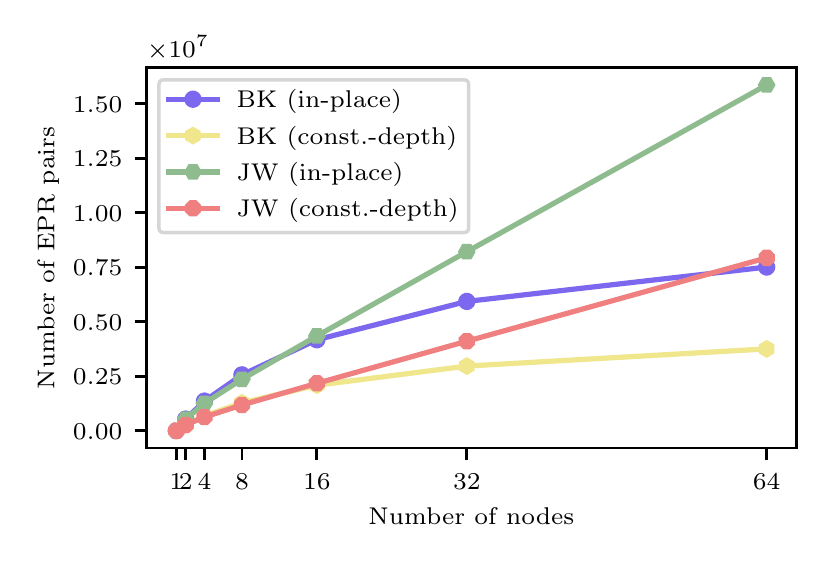}
	\caption{Number of EPR pairs required for communication to simulate one first-order Trotter step for a hydrogen ring of 32 atoms in the STO-3G basis set as a function of the number of nodes. We used either the Bravyi-Kitaev (BK) or the Jordan-Wigner (JW) encoding, see also \cref{fig:mappings}. One implementation uses the in-place circuit of \cref{fig:chem_circuits}(a) which we compare to the circuit in \cref{fig:chem_circuits}(c). The constant-depth circuit requires more local resources such as $S\geq 2$ and we additionally assumed that the rotation can be performed on an auxiliary qubit on one of the nodes already storing one of the involved orbitals. We did not consider advanced optimisations and the spin-orbitals are fixed in our example to a specific node for the full duration.}
	\label{fig:chem_comm_requirements}
\end{figure}

As the full quantum circuit is known at circuit generation time, a compiler may choose the optimal method for each term, given the available resources at that point in the program. See \cref{fig:chem_comm_requirements} for an example of a straight forward implementation without any advanced optimization applied to it.

\section{Conclusions and Outlook}

We introduce QMPI, an extension of MPI to distibuted quantum computing. This enables the development of portable high-per\-formance distributed quantum programs. Complementary, we introduce the machine-independent SENDQ performance model for distributed quantum computing. The model is motivated by technological trends in building large-scale fault-tolerant quantum machines. These considerations allowed us to simplify the model by, e.g., not modeling the overhead due to classical communication as the clock cycle rate of logical quantum operations is expected to be significantly lower. As a consequence, we end up with a deliberately simple model with only a small set of general parameters.

The SENDQ model thus allows us to expose different tradeoffs of distributed quantum algorithms in a machine-agnostic fashion and without having to deal with unnecessary details. We illustrate use cases from quantum chemistry and from condensed matter physics. Our model encourages algorithm designers to start thinking about qubit placement in a distributed setting, and overlaying communication with local computation.

The high-level modeling of the quantum network without specifying details allows hardware developers to explore implementation choices such as different quantum network topologies, and to quantify their impact in terms of the effect on the runtime of large-scale quantum computing applications.

\begin{acks}
We thank Vadym Kliuchnikov for helpful discussions. This project received support from the Microsoft Swiss Joint Research Center.
\end{acks}

\bibliographystyle{ACM-Reference-Format}
\bibliography{bibfile}


\begin{thebibliography}{65}

%%% the Web address.

\ifx \showCODEN    \undefined \def \showCODEN     #1{\unskip}     \fi
\ifx \showDOI      \undefined \def \showDOI       #1{#1}\fi
\ifx \showISBNx    \undefined \def \showISBNx     #1{\unskip}     \fi
\ifx \showISBNxiii \undefined \def \showISBNxiii  #1{\unskip}     \fi
\ifx \showISSN     \undefined \def \showISSN      #1{\unskip}     \fi
\ifx \showLCCN     \undefined \def \showLCCN      #1{\unskip}     \fi
\ifx \shownote     \undefined \def \shownote      #1{#1}          \fi
\ifx \showarticletitle \undefined \def \showarticletitle #1{#1}   \fi
\ifx \showURL      \undefined \def \showURL       {\relax}        \fi
% invisible to TeX
\providecommand\bibfield[2]{#2}
\providecommand\bibinfo[2]{#2}
\providecommand\natexlab[1]{#1}
\providecommand\showeprint[2][]{arXiv:#2}

\bibitem[\protect\citeauthoryear{Aleksandrowicz, Alexander, Barkoutsos, Bello,
  Ben-Haim, Bucher, Cabrera-Hern{\'a}ndez, Carballo-Franquis, Chen, Chen,
  et~al\mbox{.}}{Aleksandrowicz et~al\mbox{.}}{2019}]%
        {aleksandrowicz2019qiskit}
\bibfield{author}{\bibinfo{person}{Gadi Aleksandrowicz},
  \bibinfo{person}{Thomas Alexander}, \bibinfo{person}{Panagiotis Barkoutsos},
  \bibinfo{person}{Luciano Bello}, \bibinfo{person}{Yael Ben-Haim},
  \bibinfo{person}{D Bucher}, \bibinfo{person}{FJ Cabrera-Hern{\'a}ndez},
  \bibinfo{person}{J Carballo-Franquis}, \bibinfo{person}{A Chen},
  \bibinfo{person}{CF Chen}, {et~al\mbox{.}}} \bibinfo{year}{2019}\natexlab{}.
\newblock \showarticletitle{Qiskit: An open-source framework for quantum
  computing}.
\newblock \bibinfo{journal}{\emph{Accessed on: Mar}}  \bibinfo{volume}{16}
  (\bibinfo{year}{2019}).
\newblock


\bibitem[\protect\citeauthoryear{Awschalom, Berggren, Bernien, Bhave, Carr,
  Davids, Economou, Englund, Faraon, Fejer, Guha, Gustafsson, Hu, Jiang, Kim,
  Korzh, Kumar, Kwiat, Lon\ifmmode~\check{c}\else \v{c}\fi{}ar, Lukin, Miller,
  Monroe, Nam, Narang, Orcutt, Raymer, Safavi-Naeini, Spiropulu, Srinivasan,
  Sun, Vu\ifmmode \check{c}\else \v{c}\fi{}kovi\ifmmode~\acute{c}\else
  \'{c}\fi{}, Waks, Walsworth, Weiner, and Zhang}{Awschalom
  et~al\mbox{.}}{2021}]%
        {PRXQuantum.2.017002}
\bibfield{author}{\bibinfo{person}{David Awschalom}, \bibinfo{person}{Karl~K.
  Berggren}, \bibinfo{person}{Hannes Bernien}, \bibinfo{person}{Sunil Bhave},
  \bibinfo{person}{Lincoln~D. Carr}, \bibinfo{person}{Paul Davids},
  \bibinfo{person}{Sophia~E. Economou}, \bibinfo{person}{Dirk Englund},
  \bibinfo{person}{Andrei Faraon}, \bibinfo{person}{Martin Fejer},
  \bibinfo{person}{Saikat Guha}, \bibinfo{person}{Martin~V. Gustafsson},
  \bibinfo{person}{Evelyn Hu}, \bibinfo{person}{Liang Jiang},
  \bibinfo{person}{Jungsang Kim}, \bibinfo{person}{Boris Korzh},
  \bibinfo{person}{Prem Kumar}, \bibinfo{person}{Paul~G. Kwiat},
  \bibinfo{person}{Marko Lon\ifmmode~\check{c}\else \v{c}\fi{}ar},
  \bibinfo{person}{Mikhail~D. Lukin}, \bibinfo{person}{David~A.B. Miller},
  \bibinfo{person}{Christopher Monroe}, \bibinfo{person}{Sae~Woo Nam},
  \bibinfo{person}{Prineha Narang}, \bibinfo{person}{Jason~S. Orcutt},
  \bibinfo{person}{Michael~G. Raymer}, \bibinfo{person}{Amir~H. Safavi-Naeini},
  \bibinfo{person}{Maria Spiropulu}, \bibinfo{person}{Kartik Srinivasan},
  \bibinfo{person}{Shuo Sun}, \bibinfo{person}{Jelena Vu\ifmmode \check{c}\else
  \v{c}\fi{}kovi\ifmmode~\acute{c}\else \'{c}\fi{}}, \bibinfo{person}{Edo
  Waks}, \bibinfo{person}{Ronald Walsworth}, \bibinfo{person}{Andrew~M.
  Weiner}, {and} \bibinfo{person}{Zheshen Zhang}.}
  \bibinfo{year}{2021}\natexlab{}.
\newblock \showarticletitle{Development of Quantum Interconnects (QuICs) for
  Next-Generation Information Technologies}.
\newblock \bibinfo{journal}{\emph{PRX Quantum}}  \bibinfo{volume}{2}
  (\bibinfo{date}{Feb} \bibinfo{year}{2021}), \bibinfo{pages}{017002}.
\newblock
Issue 1.
\urldef\tempurl%
\url{https://doi.org/10.1103/PRXQuantum.2.017002}
\showDOI{\tempurl}


\bibitem[\protect\citeauthoryear{Beals, Brierley, Gray, Harrow, Kutin, Linden,
  Shepherd, and Stather}{Beals et~al\mbox{.}}{2013}]%
        {beals2013efficient}
\bibfield{author}{\bibinfo{person}{Robert Beals}, \bibinfo{person}{Stephen
  Brierley}, \bibinfo{person}{Oliver Gray}, \bibinfo{person}{Aram~W Harrow},
  \bibinfo{person}{Samuel Kutin}, \bibinfo{person}{Noah Linden},
  \bibinfo{person}{Dan Shepherd}, {and} \bibinfo{person}{Mark Stather}.}
  \bibinfo{year}{2013}\natexlab{}.
\newblock \showarticletitle{Efficient distributed quantum computing}.
\newblock \bibinfo{journal}{\emph{Proceedings of the Royal Society A:
  Mathematical, Physical and Engineering Sciences}} \bibinfo{volume}{469},
  \bibinfo{number}{2153} (\bibinfo{year}{2013}), \bibinfo{pages}{20120686}.
\newblock


\bibitem[\protect\citeauthoryear{Bennett}{Bennett}{1973}]%
        {bennett1973logical}
\bibfield{author}{\bibinfo{person}{Charles~H Bennett}.}
  \bibinfo{year}{1973}\natexlab{}.
\newblock \showarticletitle{Logical reversibility of computation}.
\newblock \bibinfo{journal}{\emph{IBM journal of Research and Development}}
  \bibinfo{volume}{17}, \bibinfo{number}{6} (\bibinfo{year}{1973}),
  \bibinfo{pages}{525--532}.
\newblock


\bibitem[\protect\citeauthoryear{Bennett, Brassard, Popescu, Schumacher,
  Smolin, and Wootters}{Bennett et~al\mbox{.}}{1996}]%
        {PhysRevLett.76.722}
\bibfield{author}{\bibinfo{person}{Charles~H. Bennett}, \bibinfo{person}{Gilles
  Brassard}, \bibinfo{person}{Sandu Popescu}, \bibinfo{person}{Benjamin
  Schumacher}, \bibinfo{person}{John~A. Smolin}, {and}
  \bibinfo{person}{William~K. Wootters}.} \bibinfo{year}{1996}\natexlab{}.
\newblock \showarticletitle{Purification of Noisy Entanglement and Faithful
  Teleportation via Noisy Channels}.
\newblock \bibinfo{journal}{\emph{Phys. Rev. Lett.}}  \bibinfo{volume}{76}
  (\bibinfo{date}{Jan} \bibinfo{year}{1996}), \bibinfo{pages}{722--725}.
\newblock
Issue 5.
\urldef\tempurl%
\url{https://doi.org/10.1103/PhysRevLett.76.722}
\showDOI{\tempurl}


\bibitem[\protect\citeauthoryear{Bichsel, Baader, Gehr, and Vechev}{Bichsel
  et~al\mbox{.}}{2020}]%
        {bichsel2020silq}
\bibfield{author}{\bibinfo{person}{Benjamin Bichsel},
  \bibinfo{person}{Maximilian Baader}, \bibinfo{person}{Timon Gehr}, {and}
  \bibinfo{person}{Martin Vechev}.} \bibinfo{year}{2020}\natexlab{}.
\newblock \showarticletitle{Silq: A high-level quantum language with safe
  uncomputation and intuitive semantics}. In
  \bibinfo{booktitle}{\emph{Proceedings of the 41st ACM SIGPLAN Conference on
  Programming Language Design and Implementation}}. \bibinfo{pages}{286--300}.
\newblock


\bibitem[\protect\citeauthoryear{Born and Fock}{Born and Fock}{1928}]%
        {born1928beweis}
\bibfield{author}{\bibinfo{person}{Max Born} {and} \bibinfo{person}{Vladimir
  Fock}.} \bibinfo{year}{1928}\natexlab{}.
\newblock \showarticletitle{Beweis des adiabatensatzes}.
\newblock \bibinfo{journal}{\emph{Zeitschrift f{\"u}r Physik}}
  \bibinfo{volume}{51}, \bibinfo{number}{3-4} (\bibinfo{year}{1928}),
  \bibinfo{pages}{165--180}.
\newblock


\bibitem[\protect\citeauthoryear{Bravyi and Kitaev}{Bravyi and Kitaev}{2005}]%
        {bravyi2005universal}
\bibfield{author}{\bibinfo{person}{Sergey Bravyi} {and} \bibinfo{person}{Alexei
  Kitaev}.} \bibinfo{year}{2005}\natexlab{}.
\newblock \showarticletitle{Universal quantum computation with ideal Clifford
  gates and noisy ancillas}.
\newblock \bibinfo{journal}{\emph{Physical Review A}} \bibinfo{volume}{71},
  \bibinfo{number}{2} (\bibinfo{year}{2005}), \bibinfo{pages}{022316}.
\newblock


\bibitem[\protect\citeauthoryear{Bravyi and Kitaev}{Bravyi and Kitaev}{2002}]%
        {BRAVYI2002210}
\bibfield{author}{\bibinfo{person}{Sergey~B. Bravyi} {and}
  \bibinfo{person}{Alexei~Yu. Kitaev}.} \bibinfo{year}{2002}\natexlab{}.
\newblock \showarticletitle{Fermionic Quantum Computation}.
\newblock \bibinfo{journal}{\emph{Annals of Physics}} \bibinfo{volume}{298},
  \bibinfo{number}{1} (\bibinfo{year}{2002}), \bibinfo{pages}{210--226}.
\newblock
\showISSN{0003-4916}
\urldef\tempurl%
\url{https://doi.org/10.1006/aphy.2002.6254}
\showDOI{\tempurl}


\bibitem[\protect\citeauthoryear{Collins, Linden, and Popescu}{Collins
  et~al\mbox{.}}{2001}]%
        {PhysRevA.64.032302}
\bibfield{author}{\bibinfo{person}{Daniel Collins}, \bibinfo{person}{Noah
  Linden}, {and} \bibinfo{person}{Sandu Popescu}.}
  \bibinfo{year}{2001}\natexlab{}.
\newblock \showarticletitle{Nonlocal content of quantum operations}.
\newblock \bibinfo{journal}{\emph{Phys. Rev. A}}  \bibinfo{volume}{64}
  (\bibinfo{date}{Aug} \bibinfo{year}{2001}), \bibinfo{pages}{032302}.
\newblock
Issue 3.
\urldef\tempurl%
\url{https://doi.org/10.1103/PhysRevA.64.032302}
\showDOI{\tempurl}


\bibitem[\protect\citeauthoryear{Culler, Karp, Patterson, Sahay, Schauser,
  Santos, Subramonian, and Von~Eicken}{Culler et~al\mbox{.}}{1993}]%
        {culler1993logp}
\bibfield{author}{\bibinfo{person}{David Culler}, \bibinfo{person}{Richard
  Karp}, \bibinfo{person}{David Patterson}, \bibinfo{person}{Abhijit Sahay},
  \bibinfo{person}{Klaus~Erik Schauser}, \bibinfo{person}{Eunice Santos},
  \bibinfo{person}{Ramesh Subramonian}, {and} \bibinfo{person}{Thorsten
  Von~Eicken}.} \bibinfo{year}{1993}\natexlab{}.
\newblock \showarticletitle{LogP: Towards a realistic model of parallel
  computation}. In \bibinfo{booktitle}{\emph{Proceedings of the fourth ACM
  SIGPLAN symposium on Principles and practice of parallel programming}}.
  \bibinfo{pages}{1--12}.
\newblock


\bibitem[\protect\citeauthoryear{Dahlberg, Skrzypczyk, Coopmans, Wubben,
  Rozp{\k{e}}dek, Pompili, Stolk, Pawe{\l}czak, Knegjens, de~Oliveira~Filho,
  et~al\mbox{.}}{Dahlberg et~al\mbox{.}}{2019}]%
        {dahlberg2019link}
\bibfield{author}{\bibinfo{person}{Axel Dahlberg}, \bibinfo{person}{Matthew
  Skrzypczyk}, \bibinfo{person}{Tim Coopmans}, \bibinfo{person}{Leon Wubben},
  \bibinfo{person}{Filip Rozp{\k{e}}dek}, \bibinfo{person}{Matteo Pompili},
  \bibinfo{person}{Arian Stolk}, \bibinfo{person}{Przemys{\l}aw Pawe{\l}czak},
  \bibinfo{person}{Robert Knegjens}, \bibinfo{person}{Julio de Oliveira~Filho},
  {et~al\mbox{.}}} \bibinfo{year}{2019}\natexlab{}.
\newblock \showarticletitle{A link layer protocol for quantum networks}.
\newblock In \bibinfo{booktitle}{\emph{Proceedings of the ACM Special Interest
  Group on Data Communication}}. \bibinfo{pages}{159--173}.
\newblock


\bibitem[\protect\citeauthoryear{Dahlberg and Wehner}{Dahlberg and
  Wehner}{2018}]%
        {dahlberg2018simulaqron}
\bibfield{author}{\bibinfo{person}{Axel Dahlberg} {and}
  \bibinfo{person}{Stephanie Wehner}.} \bibinfo{year}{2018}\natexlab{}.
\newblock \showarticletitle{SimulaQron—a simulator for developing quantum
  internet software}.
\newblock \bibinfo{journal}{\emph{Quantum Science and Technology}}
  \bibinfo{volume}{4}, \bibinfo{number}{1} (\bibinfo{year}{2018}),
  \bibinfo{pages}{015001}.
\newblock


\bibitem[\protect\citeauthoryear{Debone, Ouyang, Goodenough, and
  Elkouss}{Debone et~al\mbox{.}}{2020}]%
        {debone2020protocols}
\bibfield{author}{\bibinfo{person}{Sebastian Debone}, \bibinfo{person}{Runsheng
  Ouyang}, \bibinfo{person}{Kenneth Goodenough}, {and} \bibinfo{person}{David
  Elkouss}.} \bibinfo{year}{2020}\natexlab{}.
\newblock \showarticletitle{Protocols for creating and distilling multipartite
  GHZ states with Bell pairs}.
\newblock \bibinfo{journal}{\emph{IEEE Transactions on Quantum Engineering}}
  (\bibinfo{year}{2020}).
\newblock


\bibitem[\protect\citeauthoryear{DiAdamo, Ghibaudi, and Cruise}{DiAdamo
  et~al\mbox{.}}{2021}]%
        {diadamo2021distributed}
\bibfield{author}{\bibinfo{person}{Stephen DiAdamo}, \bibinfo{person}{Marco
  Ghibaudi}, {and} \bibinfo{person}{James Cruise}.}
  \bibinfo{year}{2021}\natexlab{}.
\newblock \bibinfo{title}{Distributed Quantum Computing and Network Control for
  Accelerated VQE}.
\newblock
\newblock
\showeprint[arxiv]{2101.02504}~[quant-ph]


\bibitem[\protect\citeauthoryear{Einstein, Podolsky, and Rosen}{Einstein
  et~al\mbox{.}}{1935}]%
        {PhysRev.47.777}
\bibfield{author}{\bibinfo{person}{A. Einstein}, \bibinfo{person}{B. Podolsky},
  {and} \bibinfo{person}{N. Rosen}.} \bibinfo{year}{1935}\natexlab{}.
\newblock \showarticletitle{Can Quantum-Mechanical Description of Physical
  Reality Be Considered Complete?}
\newblock \bibinfo{journal}{\emph{Phys. Rev.}}  \bibinfo{volume}{47}
  (\bibinfo{date}{May} \bibinfo{year}{1935}), \bibinfo{pages}{777--780}.
\newblock
Issue 10.
\urldef\tempurl%
\url{https://doi.org/10.1103/PhysRev.47.777}
\showDOI{\tempurl}


\bibitem[\protect\citeauthoryear{Eisert, Jacobs, Papadopoulos, and
  Plenio}{Eisert et~al\mbox{.}}{2000}]%
        {PhysRevA.62.052317}
\bibfield{author}{\bibinfo{person}{J. Eisert}, \bibinfo{person}{K. Jacobs},
  \bibinfo{person}{P. Papadopoulos}, {and} \bibinfo{person}{M.~B. Plenio}.}
  \bibinfo{year}{2000}\natexlab{}.
\newblock \showarticletitle{Optimal local implementation of nonlocal quantum
  gates}.
\newblock \bibinfo{journal}{\emph{Phys. Rev. A}}  \bibinfo{volume}{62}
  (\bibinfo{date}{Oct} \bibinfo{year}{2000}), \bibinfo{pages}{052317}.
\newblock
Issue 5.
\urldef\tempurl%
\url{https://doi.org/10.1103/PhysRevA.62.052317}
\showDOI{\tempurl}


\bibitem[\protect\citeauthoryear{Forsch, Stockill, Wallucks, Marinkovi{\'c},
  G{\"a}rtner, Norte, van Otten, Fiore, Srinivasan, and Gr{\"o}blacher}{Forsch
  et~al\mbox{.}}{2020}]%
        {forsch2020microwave}
\bibfield{author}{\bibinfo{person}{Moritz Forsch}, \bibinfo{person}{Robert
  Stockill}, \bibinfo{person}{Andreas Wallucks}, \bibinfo{person}{Igor
  Marinkovi{\'c}}, \bibinfo{person}{Claus G{\"a}rtner},
  \bibinfo{person}{Richard~A Norte}, \bibinfo{person}{Frank van Otten},
  \bibinfo{person}{Andrea Fiore}, \bibinfo{person}{Kartik Srinivasan}, {and}
  \bibinfo{person}{Simon Gr{\"o}blacher}.} \bibinfo{year}{2020}\natexlab{}.
\newblock \showarticletitle{Microwave-to-optics conversion using a mechanical
  oscillator in its quantum ground state}.
\newblock \bibinfo{journal}{\emph{Nature Physics}} \bibinfo{volume}{16},
  \bibinfo{number}{1} (\bibinfo{year}{2020}), \bibinfo{pages}{69--74}.
\newblock


\bibitem[\protect\citeauthoryear{Gambetta}{Gambetta}{2020}]%
        {IBMroadmap}
\bibfield{author}{\bibinfo{person}{Jay Gambetta}.}
  \bibinfo{year}{2020}\natexlab{}.
\newblock \bibinfo{title}{{IBM}’s Roadmap For Scaling Quantum Technology}.
\newblock
  \bibinfo{howpublished}{\url{https://www.ibm.com/blogs/research/2020/09/ibm-quantum-roadmap/}}.
\newblock
\newblock
\shownote{Accessed: 16.03.2021.}


\bibitem[\protect\citeauthoryear{Gidney and Eker{\aa}}{Gidney and
  Eker{\aa}}{2019}]%
        {gidney2019factor}
\bibfield{author}{\bibinfo{person}{Craig Gidney} {and} \bibinfo{person}{Martin
  Eker{\aa}}.} \bibinfo{year}{2019}\natexlab{}.
\newblock \showarticletitle{How to factor 2048 bit rsa integers in 8 hours
  using 20 million noisy qubits}.
\newblock \bibinfo{journal}{\emph{arXiv preprint arXiv:1905.09749}}
  (\bibinfo{year}{2019}).
\newblock


\bibitem[\protect\citeauthoryear{Green, Lumsdaine, Ross, Selinger, and
  Valiron}{Green et~al\mbox{.}}{2013}]%
        {green2013quipper}
\bibfield{author}{\bibinfo{person}{Alexander~S Green},
  \bibinfo{person}{Peter~LeFanu Lumsdaine}, \bibinfo{person}{Neil~J Ross},
  \bibinfo{person}{Peter Selinger}, {and} \bibinfo{person}{Beno{\^\i}t
  Valiron}.} \bibinfo{year}{2013}\natexlab{}.
\newblock \showarticletitle{Quipper: a scalable quantum programming language}.
  In \bibinfo{booktitle}{\emph{Proceedings of the 34th ACM SIGPLAN conference
  on Programming language design and implementation}}.
  \bibinfo{pages}{333--342}.
\newblock


\bibitem[\protect\citeauthoryear{H{\"a}ner, Jaques, Naehrig, Roetteler, and
  Soeken}{H{\"a}ner et~al\mbox{.}}{2020}]%
        {haner2020improved}
\bibfield{author}{\bibinfo{person}{Thomas H{\"a}ner}, \bibinfo{person}{Samuel
  Jaques}, \bibinfo{person}{Michael Naehrig}, \bibinfo{person}{Martin
  Roetteler}, {and} \bibinfo{person}{Mathias Soeken}.}
  \bibinfo{year}{2020}\natexlab{}.
\newblock \showarticletitle{Improved quantum circuits for elliptic curve
  discrete logarithms}. In \bibinfo{booktitle}{\emph{International Conference
  on Post-Quantum Cryptography}}. Springer, \bibinfo{pages}{425--444}.
\newblock


\bibitem[\protect\citeauthoryear{Hoefler, Kambadur, Graham, Shipman, and
  Lumsdaine}{Hoefler et~al\mbox{.}}{2007}]%
        {standard-nbc}
\bibfield{author}{\bibinfo{person}{Torsten Hoefler},
  \bibinfo{person}{Prabhanjan Kambadur}, \bibinfo{person}{Richard~L Graham},
  \bibinfo{person}{Galen Shipman}, {and} \bibinfo{person}{Andrew Lumsdaine}.}
  \bibinfo{year}{2007}\natexlab{}.
\newblock \showarticletitle{A case for standard non-blocking collective
  operations}. In \bibinfo{booktitle}{\emph{European Parallel Virtual
  Machine/Message Passing Interface Users’ Group Meeting}}. Springer,
  \bibinfo{pages}{125--134}.
\newblock


\bibitem[\protect\citeauthoryear{Hofmann, Krug, Ortegel, G{\'e}rard, Weber,
  Rosenfeld, and Weinfurter}{Hofmann et~al\mbox{.}}{2012}]%
        {Hofmann72}
\bibfield{author}{\bibinfo{person}{Julian Hofmann}, \bibinfo{person}{Michael
  Krug}, \bibinfo{person}{Norbert Ortegel}, \bibinfo{person}{Lea G{\'e}rard},
  \bibinfo{person}{Markus Weber}, \bibinfo{person}{Wenjamin Rosenfeld}, {and}
  \bibinfo{person}{Harald Weinfurter}.} \bibinfo{year}{2012}\natexlab{}.
\newblock \showarticletitle{Heralded Entanglement Between Widely Separated
  Atoms}.
\newblock \bibinfo{journal}{\emph{Science}} \bibinfo{volume}{337},
  \bibinfo{number}{6090} (\bibinfo{year}{2012}), \bibinfo{pages}{72--75}.
\newblock
\showISSN{0036-8075}
\urldef\tempurl%
\url{https://doi.org/10.1126/science.1221856}
\showDOI{\tempurl}
\showeprint{https://science.sciencemag.org/content/337/6090/72.full.pdf}


\bibitem[\protect\citeauthoryear{H{\o}yer and {\v{S}}palek}{H{\o}yer and
  {\v{S}}palek}{2005}]%
        {hoyer2005quantum}
\bibfield{author}{\bibinfo{person}{Peter H{\o}yer} {and}
  \bibinfo{person}{Robert {\v{S}}palek}.} \bibinfo{year}{2005}\natexlab{}.
\newblock \showarticletitle{Quantum fan-out is powerful}.
\newblock \bibinfo{journal}{\emph{Theory of computing}} \bibinfo{volume}{1},
  \bibinfo{number}{1} (\bibinfo{year}{2005}), \bibinfo{pages}{81--103}.
\newblock


\bibitem[\protect\citeauthoryear{JavadiAbhari, Patil, Kudrow, Heckey, Lvov,
  Chong, and Martonosi}{JavadiAbhari et~al\mbox{.}}{2015}]%
        {javadiabhari2015scaffcc}
\bibfield{author}{\bibinfo{person}{Ali JavadiAbhari}, \bibinfo{person}{Shruti
  Patil}, \bibinfo{person}{Daniel Kudrow}, \bibinfo{person}{Jeff Heckey},
  \bibinfo{person}{Alexey Lvov}, \bibinfo{person}{Frederic~T Chong}, {and}
  \bibinfo{person}{Margaret Martonosi}.} \bibinfo{year}{2015}\natexlab{}.
\newblock \showarticletitle{ScaffCC: Scalable compilation and analysis of
  quantum programs}.
\newblock \bibinfo{journal}{\emph{Parallel Comput.}}  \bibinfo{volume}{45}
  (\bibinfo{year}{2015}), \bibinfo{pages}{2--17}.
\newblock


\bibitem[\protect\citeauthoryear{Jordan and Wigner}{Jordan and Wigner}{1928}]%
        {Jordan1928}
\bibfield{author}{\bibinfo{person}{P. Jordan} {and} \bibinfo{person}{E.
  Wigner}.} \bibinfo{year}{1928}\natexlab{}.
\newblock \showarticletitle{{\"U}ber das Paulische {\"A}quivalenzverbot}.
\newblock \bibinfo{journal}{\emph{Zeitschrift f{\"u}r Physik}}
  \bibinfo{volume}{47}, \bibinfo{number}{9} (\bibinfo{year}{1928}),
  \bibinfo{pages}{631--651}.
\newblock
\urldef\tempurl%
\url{https://doi.org/10.1007/BF01331938}
\showDOI{\tempurl}


\bibitem[\protect\citeauthoryear{Kliuchnikov and Vaschillo}{Kliuchnikov and
  Vaschillo}{2021}]%
        {kliuchnikov2021}
\bibfield{author}{\bibinfo{person}{Vadym Kliuchnikov} {and}
  \bibinfo{person}{Alexander Vaschillo}.} \bibinfo{year}{2021}\natexlab{}.
\newblock \showarticletitle{Layout based on cat states}.
\newblock \bibinfo{journal}{\emph{In preparation}} (\bibinfo{year}{2021}).
\newblock


\bibitem[\protect\citeauthoryear{Lee, Berry, Gidney, Huggins, McClean, Wiebe,
  and Babbush}{Lee et~al\mbox{.}}{2020}]%
        {lee2020even}
\bibfield{author}{\bibinfo{person}{Joonho Lee}, \bibinfo{person}{Dominic
  Berry}, \bibinfo{person}{Craig Gidney}, \bibinfo{person}{William~J Huggins},
  \bibinfo{person}{Jarrod~R McClean}, \bibinfo{person}{Nathan Wiebe}, {and}
  \bibinfo{person}{Ryan Babbush}.} \bibinfo{year}{2020}\natexlab{}.
\newblock \showarticletitle{Even more efficient quantum computations of
  chemistry through tensor hypercontraction}.
\newblock \bibinfo{journal}{\emph{arXiv preprint arXiv:2011.03494}}
  (\bibinfo{year}{2020}).
\newblock


\bibitem[\protect\citeauthoryear{Lekitsch, Weidt, Fowler, M{\o}lmer, Devitt,
  Wunderlich, and Hensinger}{Lekitsch et~al\mbox{.}}{2017}]%
        {lekitsch2017blueprint}
\bibfield{author}{\bibinfo{person}{Bjoern Lekitsch}, \bibinfo{person}{Sebastian
  Weidt}, \bibinfo{person}{Austin~G Fowler}, \bibinfo{person}{Klaus M{\o}lmer},
  \bibinfo{person}{Simon~J Devitt}, \bibinfo{person}{Christof Wunderlich},
  {and} \bibinfo{person}{Winfried~K Hensinger}.}
  \bibinfo{year}{2017}\natexlab{}.
\newblock \showarticletitle{Blueprint for a microwave trapped ion quantum
  computer}.
\newblock \bibinfo{journal}{\emph{Science Advances}} \bibinfo{volume}{3},
  \bibinfo{number}{2} (\bibinfo{year}{2017}), \bibinfo{pages}{e1601540}.
\newblock


\bibitem[\protect\citeauthoryear{Litinski}{Litinski}{2019}]%
        {litinski2019magic}
\bibfield{author}{\bibinfo{person}{Daniel Litinski}.}
  \bibinfo{year}{2019}\natexlab{}.
\newblock \showarticletitle{Magic state distillation: Not as costly as you
  think}.
\newblock \bibinfo{journal}{\emph{Quantum}}  \bibinfo{volume}{3}
  (\bibinfo{year}{2019}), \bibinfo{pages}{205}.
\newblock


\bibitem[\protect\citeauthoryear{Magnard, Storz, Kurpiers, Sch{\"a}r, Marxer,
  L{\"u}tolf, Walter, Besse, Gabureac, Reuer, et~al\mbox{.}}{Magnard
  et~al\mbox{.}}{2020}]%
        {magnard2020microwave}
\bibfield{author}{\bibinfo{person}{Paul Magnard}, \bibinfo{person}{Simon
  Storz}, \bibinfo{person}{Philipp Kurpiers}, \bibinfo{person}{Josua
  Sch{\"a}r}, \bibinfo{person}{Fabian Marxer}, \bibinfo{person}{Janis
  L{\"u}tolf}, \bibinfo{person}{T Walter}, \bibinfo{person}{J-C Besse},
  \bibinfo{person}{M Gabureac}, \bibinfo{person}{K Reuer}, {et~al\mbox{.}}}
  \bibinfo{year}{2020}\natexlab{}.
\newblock \showarticletitle{Microwave quantum link between superconducting
  circuits housed in spatially separated cryogenic systems}.
\newblock \bibinfo{journal}{\emph{Physical Review Letters}}
  \bibinfo{volume}{125}, \bibinfo{number}{26} (\bibinfo{year}{2020}),
  \bibinfo{pages}{260502}.
\newblock


\bibitem[\protect\citeauthoryear{McArdle, Endo, Aspuru-Guzik, Benjamin, and
  Yuan}{McArdle et~al\mbox{.}}{2020}]%
        {RevModPhys.92.015003}
\bibfield{author}{\bibinfo{person}{Sam McArdle}, \bibinfo{person}{Suguru Endo},
  \bibinfo{person}{Al\'an Aspuru-Guzik}, \bibinfo{person}{Simon~C. Benjamin},
  {and} \bibinfo{person}{Xiao Yuan}.} \bibinfo{year}{2020}\natexlab{}.
\newblock \showarticletitle{Quantum computational chemistry}.
\newblock \bibinfo{journal}{\emph{Rev. Mod. Phys.}}  \bibinfo{volume}{92}
  (\bibinfo{date}{Mar} \bibinfo{year}{2020}), \bibinfo{pages}{015003}.
\newblock
Issue 1.
\urldef\tempurl%
\url{https://doi.org/10.1103/RevModPhys.92.015003}
\showDOI{\tempurl}


\bibitem[\protect\citeauthoryear{McClean, Rubin, Sung, Kivlichan,
  Bonet-Monroig, Cao, Dai, Fried, Gidney, Gimby, et~al\mbox{.}}{McClean
  et~al\mbox{.}}{2020}]%
        {mcclean2020}
\bibfield{author}{\bibinfo{person}{Jarrod~R McClean},
  \bibinfo{person}{Nicholas~C Rubin}, \bibinfo{person}{Kevin~J Sung},
  \bibinfo{person}{Ian~D Kivlichan}, \bibinfo{person}{Xavier Bonet-Monroig},
  \bibinfo{person}{Yudong Cao}, \bibinfo{person}{Chengyu Dai},
  \bibinfo{person}{E~Schuyler Fried}, \bibinfo{person}{Craig Gidney},
  \bibinfo{person}{Brendan Gimby}, {et~al\mbox{.}}}
  \bibinfo{year}{2020}\natexlab{}.
\newblock \showarticletitle{OpenFermion: the electronic structure package for
  quantum computers}.
\newblock \bibinfo{journal}{\emph{Quantum Science and Technology}}
  \bibinfo{volume}{5}, \bibinfo{number}{3} (\bibinfo{year}{2020}),
  \bibinfo{pages}{034014}.
\newblock


\bibitem[\protect\citeauthoryear{Meter, Munro, Nemoto, and Itoh}{Meter
  et~al\mbox{.}}{2008}]%
        {meter2008arithmetic}
\bibfield{author}{\bibinfo{person}{Rodney~Van Meter}, \bibinfo{person}{WJ
  Munro}, \bibinfo{person}{Kae Nemoto}, {and} \bibinfo{person}{Kohei~M Itoh}.}
  \bibinfo{year}{2008}\natexlab{}.
\newblock \showarticletitle{Arithmetic on a distributed-memory quantum
  multicomputer}.
\newblock \bibinfo{journal}{\emph{ACM Journal on Emerging Technologies in
  Computing Systems (JETC)}} \bibinfo{volume}{3}, \bibinfo{number}{4}
  (\bibinfo{year}{2008}), \bibinfo{pages}{1--23}.
\newblock


\bibitem[\protect\citeauthoryear{Meter~III}{Meter~III}{2006}]%
        {meter2006architecture}
\bibfield{author}{\bibinfo{person}{Rodney Doyle~Van Meter~III}.}
  \bibinfo{year}{2006}\natexlab{}.
\newblock \showarticletitle{Architecture of a quantum multicomputer optimized
  for shor's factoring algorithm}.
\newblock \bibinfo{journal}{\emph{arXiv preprint quant-ph/0607065}}
  (\bibinfo{year}{2006}).
\newblock


\bibitem[\protect\citeauthoryear{Moehring, Maunz, Olmschenk, Younge,
  Matsukevich, Duan, and Monroe}{Moehring et~al\mbox{.}}{2007}]%
        {moehring2007entanglement}
\bibfield{author}{\bibinfo{person}{David~L Moehring}, \bibinfo{person}{Peter
  Maunz}, \bibinfo{person}{Steve Olmschenk}, \bibinfo{person}{Kelly~C Younge},
  \bibinfo{person}{Dzmitry~N Matsukevich}, \bibinfo{person}{L-M Duan}, {and}
  \bibinfo{person}{Christopher Monroe}.} \bibinfo{year}{2007}\natexlab{}.
\newblock \showarticletitle{Entanglement of single-atom quantum bits at a
  distance}.
\newblock \bibinfo{journal}{\emph{Nature}} \bibinfo{volume}{449},
  \bibinfo{number}{7158} (\bibinfo{year}{2007}), \bibinfo{pages}{68--71}.
\newblock


\bibitem[\protect\citeauthoryear{Monroe, Raussendorf, Ruthven, Brown, Maunz,
  Duan, and Kim}{Monroe et~al\mbox{.}}{2014}]%
        {PhysRevA.89.022317}
\bibfield{author}{\bibinfo{person}{C. Monroe}, \bibinfo{person}{R.
  Raussendorf}, \bibinfo{person}{A. Ruthven}, \bibinfo{person}{K.~R. Brown},
  \bibinfo{person}{P. Maunz}, \bibinfo{person}{L.-M. Duan}, {and}
  \bibinfo{person}{J. Kim}.} \bibinfo{year}{2014}\natexlab{}.
\newblock \showarticletitle{Large-scale modular quantum-computer architecture
  with atomic memory and photonic interconnects}.
\newblock \bibinfo{journal}{\emph{Phys. Rev. A}}  \bibinfo{volume}{89}
  (\bibinfo{date}{Feb} \bibinfo{year}{2014}), \bibinfo{pages}{022317}.
\newblock
Issue 2.
\urldef\tempurl%
\url{https://doi.org/10.1103/PhysRevA.89.022317}
\showDOI{\tempurl}


\bibitem[\protect\citeauthoryear{Neven}{Neven}{2020}]%
        {googleroadmap}
\bibfield{author}{\bibinfo{person}{Hartmut Neven}.}
  \bibinfo{year}{2020}\natexlab{}.
\newblock \bibinfo{title}{Google Quantum AI updates at Quantum Summer Symposium
  2020}.
\newblock
\newblock
\urldef\tempurl%
\url{{https://www.youtube.com/watch?v=TJ6vBNEQReU}}
\showURL{%
\tempurl}
\newblock
\shownote{Online; posted 3-September-2020, accessed 25-March-2021.}


\bibitem[\protect\citeauthoryear{Nickerson, Fitzsimons, and Benjamin}{Nickerson
  et~al\mbox{.}}{2014}]%
        {PhysRevX.4.041041}
\bibfield{author}{\bibinfo{person}{Naomi~H. Nickerson},
  \bibinfo{person}{Joseph~F. Fitzsimons}, {and} \bibinfo{person}{Simon~C.
  Benjamin}.} \bibinfo{year}{2014}\natexlab{}.
\newblock \showarticletitle{Freely Scalable Quantum Technologies Using Cells of
  5-to-50 Qubits with Very Lossy and Noisy Photonic Links}.
\newblock \bibinfo{journal}{\emph{Phys. Rev. X}}  \bibinfo{volume}{4}
  (\bibinfo{date}{Dec} \bibinfo{year}{2014}), \bibinfo{pages}{041041}.
\newblock
Issue 4.
\urldef\tempurl%
\url{https://doi.org/10.1103/PhysRevX.4.041041}
\showDOI{\tempurl}


\bibitem[\protect\citeauthoryear{Nielsen and Chuang}{Nielsen and
  Chuang}{2002}]%
        {nielsen2002quantum}
\bibfield{author}{\bibinfo{person}{Michael~A Nielsen} {and}
  \bibinfo{person}{Isaac Chuang}.} \bibinfo{year}{2002}\natexlab{}.
\newblock \bibinfo{title}{Quantum computation and quantum information}.
\newblock
\newblock


\bibitem[\protect\citeauthoryear{Ortiz, Gubernatis, Knill, and Laflamme}{Ortiz
  et~al\mbox{.}}{2001}]%
        {ortiz2001quantum}
\bibfield{author}{\bibinfo{person}{Gerardo Ortiz}, \bibinfo{person}{James~E
  Gubernatis}, \bibinfo{person}{Emanuel Knill}, {and} \bibinfo{person}{Raymond
  Laflamme}.} \bibinfo{year}{2001}\natexlab{}.
\newblock \showarticletitle{Quantum algorithms for fermionic simulations}.
\newblock \bibinfo{journal}{\emph{Physical Review A}} \bibinfo{volume}{64},
  \bibinfo{number}{2} (\bibinfo{year}{2001}), \bibinfo{pages}{022319}.
\newblock


\bibitem[\protect\citeauthoryear{Pirandola, Andersen, Banchi, Berta, Bunandar,
  Colbeck, Englund, Gehring, Lupo, Ottaviani, et~al\mbox{.}}{Pirandola
  et~al\mbox{.}}{2020}]%
        {pirandola2020advances}
\bibfield{author}{\bibinfo{person}{Stefano Pirandola}, \bibinfo{person}{Ulrik~L
  Andersen}, \bibinfo{person}{Leonardo Banchi}, \bibinfo{person}{Mario Berta},
  \bibinfo{person}{Darius Bunandar}, \bibinfo{person}{Roger Colbeck},
  \bibinfo{person}{Dirk Englund}, \bibinfo{person}{Tobias Gehring},
  \bibinfo{person}{Cosmo Lupo}, \bibinfo{person}{Carlo Ottaviani},
  {et~al\mbox{.}}} \bibinfo{year}{2020}\natexlab{}.
\newblock \showarticletitle{Advances in quantum cryptography}.
\newblock \bibinfo{journal}{\emph{Advances in Optics and Photonics}}
  \bibinfo{volume}{12}, \bibinfo{number}{4} (\bibinfo{year}{2020}),
  \bibinfo{pages}{1012--1236}.
\newblock


\bibitem[\protect\citeauthoryear{Reiher, Wiebe, Svore, Wecker, and
  Troyer}{Reiher et~al\mbox{.}}{2017}]%
        {reiher2017elucidating}
\bibfield{author}{\bibinfo{person}{Markus Reiher}, \bibinfo{person}{Nathan
  Wiebe}, \bibinfo{person}{Krysta~M Svore}, \bibinfo{person}{Dave Wecker},
  {and} \bibinfo{person}{Matthias Troyer}.} \bibinfo{year}{2017}\natexlab{}.
\newblock \showarticletitle{Elucidating reaction mechanisms on quantum
  computers}.
\newblock \bibinfo{journal}{\emph{Proceedings of the National Academy of
  Sciences}} \bibinfo{volume}{114}, \bibinfo{number}{29}
  (\bibinfo{year}{2017}), \bibinfo{pages}{7555--7560}.
\newblock


\bibitem[\protect\citeauthoryear{Sanders and Tr{\"a}ff}{Sanders and
  Tr{\"a}ff}{2006}]%
        {sanders2006parallel}
\bibfield{author}{\bibinfo{person}{Peter Sanders} {and}
  \bibinfo{person}{Jesper~Larsson Tr{\"a}ff}.} \bibinfo{year}{2006}\natexlab{}.
\newblock \showarticletitle{Parallel prefix (scan) algorithms for MPI}. In
  \bibinfo{booktitle}{\emph{European Parallel Virtual Machine/Message Passing
  Interface Users’ Group Meeting}}. Springer, \bibinfo{pages}{49--57}.
\newblock


\bibitem[\protect\citeauthoryear{Sch{\"a}fer, Ballance, Thirumalai, Stephenson,
  Ballance, Steane, and Lucas}{Sch{\"a}fer et~al\mbox{.}}{2018}]%
        {schafer2018fast}
\bibfield{author}{\bibinfo{person}{VM Sch{\"a}fer}, \bibinfo{person}{CJ
  Ballance}, \bibinfo{person}{K Thirumalai}, \bibinfo{person}{LJ Stephenson},
  \bibinfo{person}{TG Ballance}, \bibinfo{person}{AM Steane}, {and}
  \bibinfo{person}{DM Lucas}.} \bibinfo{year}{2018}\natexlab{}.
\newblock \showarticletitle{Fast quantum logic gates with trapped-ion qubits}.
\newblock \bibinfo{journal}{\emph{Nature}} \bibinfo{volume}{555},
  \bibinfo{number}{7694} (\bibinfo{year}{2018}), \bibinfo{pages}{75--78}.
\newblock


\bibitem[\protect\citeauthoryear{Scherer, Valiron, Mau, Alexander, Van~den
  Berg, and Chapuran}{Scherer et~al\mbox{.}}{2017}]%
        {scherer2017concrete}
\bibfield{author}{\bibinfo{person}{Artur Scherer}, \bibinfo{person}{Beno{\^\i}t
  Valiron}, \bibinfo{person}{Siun-Chuon Mau}, \bibinfo{person}{Scott
  Alexander}, \bibinfo{person}{Eric Van~den Berg}, {and}
  \bibinfo{person}{Thomas~E Chapuran}.} \bibinfo{year}{2017}\natexlab{}.
\newblock \showarticletitle{Concrete resource analysis of the quantum
  linear-system algorithm used to compute the electromagnetic scattering cross
  section of a 2D target}.
\newblock \bibinfo{journal}{\emph{Quantum Information Processing}}
  \bibinfo{volume}{16}, \bibinfo{number}{3} (\bibinfo{year}{2017}),
  \bibinfo{pages}{1--65}.
\newblock


\bibitem[\protect\citeauthoryear{Shor}{Shor}{1994}]%
        {shor1994algorithms}
\bibfield{author}{\bibinfo{person}{Peter~W Shor}.}
  \bibinfo{year}{1994}\natexlab{}.
\newblock \showarticletitle{Algorithms for quantum computation: discrete
  logarithms and factoring}. In \bibinfo{booktitle}{\emph{Proceedings 35th
  annual symposium on foundations of computer science}}. Ieee,
  \bibinfo{pages}{124--134}.
\newblock


\bibitem[\protect\citeauthoryear{Somma, Ortiz, Gubernatis, Knill, and
  Laflamme}{Somma et~al\mbox{.}}{2002}]%
        {somma2002simulating}
\bibfield{author}{\bibinfo{person}{Rolando Somma}, \bibinfo{person}{Gerardo
  Ortiz}, \bibinfo{person}{James~E Gubernatis}, \bibinfo{person}{Emanuel
  Knill}, {and} \bibinfo{person}{Raymond Laflamme}.}
  \bibinfo{year}{2002}\natexlab{}.
\newblock \showarticletitle{Simulating physical phenomena by quantum networks}.
\newblock \bibinfo{journal}{\emph{Physical Review A}} \bibinfo{volume}{65},
  \bibinfo{number}{4} (\bibinfo{year}{2002}), \bibinfo{pages}{042323}.
\newblock


\bibitem[\protect\citeauthoryear{Steiger, H{\"a}ner, and Troyer}{Steiger
  et~al\mbox{.}}{2018}]%
        {steiger2018projectq}
\bibfield{author}{\bibinfo{person}{Damian~S Steiger}, \bibinfo{person}{Thomas
  H{\"a}ner}, {and} \bibinfo{person}{Matthias Troyer}.}
  \bibinfo{year}{2018}\natexlab{}.
\newblock \showarticletitle{ProjectQ: an open source software framework for
  quantum computing}.
\newblock \bibinfo{journal}{\emph{Quantum}}  \bibinfo{volume}{2}
  (\bibinfo{year}{2018}), \bibinfo{pages}{49}.
\newblock


\bibitem[\protect\citeauthoryear{Sun, Berkelbach, Blunt, Booth, Guo, Li, Liu,
  McClain, Sayfutyarova, Sharma, Wouters, and Chan}{Sun et~al\mbox{.}}{2017}]%
        {PYSCF}
\bibfield{author}{\bibinfo{person}{Qiming Sun}, \bibinfo{person}{Timothy~C.
  Berkelbach}, \bibinfo{person}{Nick~S. Blunt}, \bibinfo{person}{George~H.
  Booth}, \bibinfo{person}{Sheng Guo}, \bibinfo{person}{Zhendong Li},
  \bibinfo{person}{Junzi Liu}, \bibinfo{person}{James~D. McClain},
  \bibinfo{person}{Elvira~R. Sayfutyarova}, \bibinfo{person}{Sandeep Sharma},
  \bibinfo{person}{Sebastian Wouters}, {and} \bibinfo{person}{Garnet~Kin‐Lic
  Chan}.} \bibinfo{year}{2017}\natexlab{}.
\newblock \bibinfo{title}{PySCF: the Python‐based simulations of chemistry
  framework}.
\newblock , \bibinfo{numpages}{e1340}~pages.
\newblock
\urldef\tempurl%
\url{https://doi.org/10.1002/wcms.1340}
\showDOI{\tempurl}
\showeprint{https://onlinelibrary.wiley.com/doi/pdf/10.1002/wcms.1340}


\bibitem[\protect\citeauthoryear{Svore, Geller, Troyer, Azariah, Granade, Heim,
  Kliuchnikov, Mykhailova, Paz, and Roetteler}{Svore et~al\mbox{.}}{2018}]%
        {svore2018q}
\bibfield{author}{\bibinfo{person}{Krysta Svore}, \bibinfo{person}{Alan
  Geller}, \bibinfo{person}{Matthias Troyer}, \bibinfo{person}{John Azariah},
  \bibinfo{person}{Christopher Granade}, \bibinfo{person}{Bettina Heim},
  \bibinfo{person}{Vadym Kliuchnikov}, \bibinfo{person}{Mariia Mykhailova},
  \bibinfo{person}{Andres Paz}, {and} \bibinfo{person}{Martin Roetteler}.}
  \bibinfo{year}{2018}\natexlab{}.
\newblock \showarticletitle{Q\# enabling scalable quantum computing and
  development with a high-level dsl}. In \bibinfo{booktitle}{\emph{Proceedings
  of the Real World Domain Specific Languages Workshop 2018}}.
  \bibinfo{pages}{1--10}.
\newblock


\bibitem[\protect\citeauthoryear{Tsuchimoto, Kn\"uppel, Delteil, Sun, Kroner,
  and Imamo\ifmmode~\breve{g}\else \u{g}\fi{}lu}{Tsuchimoto
  et~al\mbox{.}}{2017}]%
        {PhysRevB.96.165312}
\bibfield{author}{\bibinfo{person}{Yuta Tsuchimoto}, \bibinfo{person}{Patrick
  Kn\"uppel}, \bibinfo{person}{Aymeric Delteil}, \bibinfo{person}{Zhe Sun},
  \bibinfo{person}{Martin Kroner}, {and} \bibinfo{person}{Ata\ifmmode
  \mbox{\c{c}}\else~\c{c}\fi{} Imamo\ifmmode~\breve{g}\else \u{g}\fi{}lu}.}
  \bibinfo{year}{2017}\natexlab{}.
\newblock \showarticletitle{Proposal for a quantum interface between photonic
  and superconducting qubits}.
\newblock \bibinfo{journal}{\emph{Phys. Rev. B}}  \bibinfo{volume}{96}
  (\bibinfo{date}{Oct} \bibinfo{year}{2017}), \bibinfo{pages}{165312}.
\newblock
Issue 16.
\urldef\tempurl%
\url{https://doi.org/10.1103/PhysRevB.96.165312}
\showDOI{\tempurl}


\bibitem[\protect\citeauthoryear{van Enk, Cirac, and Zoller}{van Enk
  et~al\mbox{.}}{1997}]%
        {PhysRevLett.78.4293}
\bibfield{author}{\bibinfo{person}{S.~J. van Enk}, \bibinfo{person}{J.~I.
  Cirac}, {and} \bibinfo{person}{P. Zoller}.} \bibinfo{year}{1997}\natexlab{}.
\newblock \showarticletitle{Ideal Quantum Communication over Noisy Channels: A
  Quantum Optical Implementation}.
\newblock \bibinfo{journal}{\emph{Phys. Rev. Lett.}}  \bibinfo{volume}{78}
  (\bibinfo{date}{Jun} \bibinfo{year}{1997}), \bibinfo{pages}{4293--4296}.
\newblock
Issue 22.
\urldef\tempurl%
\url{https://doi.org/10.1103/PhysRevLett.78.4293}
\showDOI{\tempurl}


\bibitem[\protect\citeauthoryear{Van~Meter and Devitt}{Van~Meter and
  Devitt}{2016}]%
        {van2016path}
\bibfield{author}{\bibinfo{person}{Rodney Van~Meter} {and}
  \bibinfo{person}{Simon~J Devitt}.} \bibinfo{year}{2016}\natexlab{}.
\newblock \showarticletitle{The path to scalable distributed quantum
  computing}.
\newblock \bibinfo{journal}{\emph{Computer}} \bibinfo{volume}{49},
  \bibinfo{number}{9} (\bibinfo{year}{2016}), \bibinfo{pages}{31--42}.
\newblock


\bibitem[\protect\citeauthoryear{Van~Meter, Nemoto, and Munro}{Van~Meter
  et~al\mbox{.}}{2007}]%
        {van2007communication}
\bibfield{author}{\bibinfo{person}{Rod Van~Meter}, \bibinfo{person}{Kae
  Nemoto}, {and} \bibinfo{person}{W Munro}.} \bibinfo{year}{2007}\natexlab{}.
\newblock \showarticletitle{Communication links for distributed quantum
  computation}.
\newblock \bibinfo{journal}{\emph{IEEE Trans. Comput.}} \bibinfo{volume}{56},
  \bibinfo{number}{12} (\bibinfo{year}{2007}), \bibinfo{pages}{1643--1653}.
\newblock


\bibitem[\protect\citeauthoryear{von Burg, Low, H{\"a}ner, Steiger, Reiher,
  Roetteler, and Troyer}{von Burg et~al\mbox{.}}{2020}]%
        {von2020quantum}
\bibfield{author}{\bibinfo{person}{Vera von Burg}, \bibinfo{person}{Guang~Hao
  Low}, \bibinfo{person}{Thomas H{\"a}ner}, \bibinfo{person}{Damian~S Steiger},
  \bibinfo{person}{Markus Reiher}, \bibinfo{person}{Martin Roetteler}, {and}
  \bibinfo{person}{Matthias Troyer}.} \bibinfo{year}{2020}\natexlab{}.
\newblock \showarticletitle{Quantum computing enhanced computational
  catalysis}.
\newblock \bibinfo{journal}{\emph{arXiv preprint arXiv:2007.14460}}
  (\bibinfo{year}{2020}).
\newblock


\bibitem[\protect\citeauthoryear{Walker and Dongarra}{Walker and
  Dongarra}{1996}]%
        {walker1996mpi}
\bibfield{author}{\bibinfo{person}{David~W Walker} {and}
  \bibinfo{person}{Jack~J Dongarra}.} \bibinfo{year}{1996}\natexlab{}.
\newblock \showarticletitle{MPI: a standard message passing interface}.
\newblock \bibinfo{journal}{\emph{Supercomputer}}  \bibinfo{volume}{12}
  (\bibinfo{year}{1996}), \bibinfo{pages}{56--68}.
\newblock


\bibitem[\protect\citeauthoryear{Watts, Kothari, Schaeffer, and Tal}{Watts
  et~al\mbox{.}}{2019}]%
        {10.1145/3313276.3316404}
\bibfield{author}{\bibinfo{person}{Adam~Bene Watts}, \bibinfo{person}{Robin
  Kothari}, \bibinfo{person}{Luke Schaeffer}, {and} \bibinfo{person}{Avishay
  Tal}.} \bibinfo{year}{2019}\natexlab{}.
\newblock \showarticletitle{Exponential Separation between Shallow Quantum
  Circuits and Unbounded Fan-in Shallow Classical Circuits}. In
  \bibinfo{booktitle}{\emph{Proceedings of the 51st Annual ACM SIGACT Symposium
  on Theory of Computing}} (Phoenix, AZ, USA) \emph{(\bibinfo{series}{STOC
  2019})}. \bibinfo{publisher}{Association for Computing Machinery},
  \bibinfo{address}{New York, NY, USA}, \bibinfo{pages}{515–526}.
\newblock
\showISBNx{9781450367059}
\urldef\tempurl%
\url{https://doi.org/10.1145/3313276.3316404}
\showDOI{\tempurl}


\bibitem[\protect\citeauthoryear{Wehner, Elkouss, and Hanson}{Wehner
  et~al\mbox{.}}{2018}]%
        {wehner2018quantum}
\bibfield{author}{\bibinfo{person}{Stephanie Wehner}, \bibinfo{person}{David
  Elkouss}, {and} \bibinfo{person}{Ronald Hanson}.}
  \bibinfo{year}{2018}\natexlab{}.
\newblock \showarticletitle{Quantum internet: A vision for the road ahead}.
\newblock \bibinfo{journal}{\emph{Science}} \bibinfo{volume}{362},
  \bibinfo{number}{6412} (\bibinfo{year}{2018}).
\newblock


\bibitem[\protect\citeauthoryear{Whitfield, Havl\'{\i}\ifmmode~\check{c}\else
  \v{c}\fi{}ek, and Troyer}{Whitfield et~al\mbox{.}}{2016}]%
        {PhysRevA.94.030301}
\bibfield{author}{\bibinfo{person}{James~D. Whitfield},
  \bibinfo{person}{Vojt\ifmmode \check{e}\else~\v{e}\fi{}ch
  Havl\'{\i}\ifmmode~\check{c}\else \v{c}\fi{}ek}, {and}
  \bibinfo{person}{Matthias Troyer}.} \bibinfo{year}{2016}\natexlab{}.
\newblock \showarticletitle{Local spin operators for fermion simulations}.
\newblock \bibinfo{journal}{\emph{Phys. Rev. A}}  \bibinfo{volume}{94}
  (\bibinfo{date}{Sep} \bibinfo{year}{2016}), \bibinfo{pages}{030301}.
\newblock
Issue 3.
\urldef\tempurl%
\url{https://doi.org/10.1103/PhysRevA.94.030301}
\showDOI{\tempurl}


\bibitem[\protect\citeauthoryear{Xiang, Zhang, Jiang, and Rabl}{Xiang
  et~al\mbox{.}}{2017}]%
        {xiang2017intracity}
\bibfield{author}{\bibinfo{person}{Ze-Liang Xiang}, \bibinfo{person}{Mengzhen
  Zhang}, \bibinfo{person}{Liang Jiang}, {and} \bibinfo{person}{Peter Rabl}.}
  \bibinfo{year}{2017}\natexlab{}.
\newblock \showarticletitle{Intracity quantum communication via thermal
  microwave networks}.
\newblock \bibinfo{journal}{\emph{Physical Review X}} \bibinfo{volume}{7},
  \bibinfo{number}{1} (\bibinfo{year}{2017}), \bibinfo{pages}{011035}.
\newblock


\bibitem[\protect\citeauthoryear{Yimsiriwattana and Lomonaco~Jr}{Yimsiriwattana
  and Lomonaco~Jr}{2004a}]%
        {yimsiriwattana2004distributed}
\bibfield{author}{\bibinfo{person}{Anocha Yimsiriwattana} {and}
  \bibinfo{person}{Samuel~J Lomonaco~Jr}.} \bibinfo{year}{2004}\natexlab{a}.
\newblock \showarticletitle{Distributed quantum computing: A distributed Shor
  algorithm}. In \bibinfo{booktitle}{\emph{Quantum Information and Computation
  II}}, Vol.~\bibinfo{volume}{5436}. International Society for Optics and
  Photonics, \bibinfo{pages}{360--372}.
\newblock


\bibitem[\protect\citeauthoryear{Yimsiriwattana and Lomonaco~Jr}{Yimsiriwattana
  and Lomonaco~Jr}{2004b}]%
        {yimsiriwattana2004generalized}
\bibfield{author}{\bibinfo{person}{Anocha Yimsiriwattana} {and}
  \bibinfo{person}{Samuel~J Lomonaco~Jr}.} \bibinfo{year}{2004}\natexlab{b}.
\newblock \showarticletitle{Generalized GHZ states and distributed quantum
  computing}.
\newblock \bibinfo{journal}{\emph{arXiv preprint quant-ph/0402148}}
  (\bibinfo{year}{2004}).
\newblock


\bibitem[\protect\citeauthoryear{Zhong, Chang, Bienfait, Dumur, Chou, Conner,
  Grebel, Povey, Yan, Schuster, et~al\mbox{.}}{Zhong et~al\mbox{.}}{2021}]%
        {zhong2021deterministic}
\bibfield{author}{\bibinfo{person}{Youpeng Zhong}, \bibinfo{person}{Hung-Shen
  Chang}, \bibinfo{person}{Audrey Bienfait}, \bibinfo{person}{{\'E}tienne
  Dumur}, \bibinfo{person}{Ming-Han Chou}, \bibinfo{person}{Christopher~R
  Conner}, \bibinfo{person}{Joel Grebel}, \bibinfo{person}{Rhys~G Povey},
  \bibinfo{person}{Haoxiong Yan}, \bibinfo{person}{David~I Schuster},
  {et~al\mbox{.}}} \bibinfo{year}{2021}\natexlab{}.
\newblock \showarticletitle{Deterministic multi-qubit entanglement in a quantum
  network}.
\newblock \bibinfo{journal}{\emph{Nature}} \bibinfo{volume}{590},
  \bibinfo{number}{7847} (\bibinfo{year}{2021}), \bibinfo{pages}{571--575}.
\newblock


\end{thebibliography}

\appendix

\section{Example Implementations}

\subsection{Moving Qubits} Here, we give an example implementation of \qmpi{Send\_move} and \qmpi{Recv\_move} in our QMPI prototype. The sender executes \qmpi{Send\_move}, which sends a qubit (with move semantics), and the receiver calls the corresponding \qmpi{Recv\_move}. Both of these functions can be implemented using EPR-pair preparation and local quantum operations as follows:
\begin{lstlisting}[style=cppstyle,escapechar=\%]
void %\qmpi{Send\_move}%(QMPI_QUBIT_PTR qubit, int dest, int tag, MPI_Comm comm) {
	auto epr_qubit = QMPI_Alloc_qmem(1);
	QMPI_Prepare_EPR(epr_qubit, dest, tag, comm);
	CNOT(qubit, epr_qubit);
	int r=0;
	r = Measure(epr_qubit);
	H(qubit);
	r |= 2 * Measure(qubit);
	QMPI_Free_qmem(epr_qubit, 1);
	MPI_Send(&r, 1, MPI_INT, dest, tag, comm);
}
void %\qmpi{Recv\_move}%(QMPI_QUBIT_PTR qubit, int src, int tag, MPI_Comm comm) {
	QMPI_Prepare_EPR(qubit, src, tag, comm);
	int r;
	MPI_Recv(&r, 1, MPI_INT, src, tag, comm, MPI_STATUS_IGNORE);
	if (r&1)
		X(qubit);
	if (r&2)
		Z(qubit);
}
\end{lstlisting}

We note that these functions may also be implemented by relying on \qmpi{Send / Recv} and their inverses: Once the value of a qubit is shared between two nodes, it is no longer possible to distinguish sender from receiver. Therefore, the two involved nodes may exchange roles when calling the inverses of \qmpi{Send / Recv}, resulting in a slightly less efficient implementation of teleportation (since measurement results are communicated using two one-bit messages instead of one two-bit message).

\subsection{Transverse-field Ising Model (TFIM).} As a second code example, we provide an implementation of time evolution under a TFIM Hamiltonian below. Note that the code also includes annealing from a fully transverse-field model to a fully classical Ising model.
We note that the code can be significantly optimized by using asynchronous communication primitives. However, we use blocking calls only to simplify the presentation.

\begin{lstlisting}[style=cppstyle,caption={QMPI code for TFIM time evolution and annealing.},label={lst:tfimcode}]
#include "qmpi.hpp"
#include <iostream>

using namespace QMPI;

void tfim_time_evolution(double const& J, double const& g, double const& time, QMPI_QUBIT_PTR qubits, unsigned num_spins, unsigned num_trotter) {
	int rank, size;
	QMPI_Comm_size(QMPI_COMM_WORLD, &size);
	QMPI_Comm_rank(QMPI_COMM_WORLD, &rank);
	auto dt = time/num_trotter;
	for (unsigned step=0; step < num_trotter; ++step) {
		for (unsigned site = 0; site < num_spins-1; ++site) {
			CNOT(qubits+site, qubits+site+1);
			Rz(qubits+site+1, 2.0 * J * dt);
			CNOT(qubits+site, qubits+site+1);
		}
		if (size == 1) { // single rank: no communication required
			CNOT(qubits+num_spins-1, qubits);
			Rz(qubits, 2.0 * J * dt);
			CNOT(qubits+num_spins-1, qubits);
		}
		else {
			for (unsigned odd = 0; odd < 2; ++odd) {
				if ((rank&1) == odd) {
					QMPI_Send(qubits, (rank-1+size)%size, 0, QMPI_COMM_WORLD);
					QMPI_Unsend(qubits, (rank-1+size)%size, 0, QMPI_COMM_WORLD);
				}
				else {
					auto tmpqubit = QMPI_Alloc_qmem(1);
					QMPI_Recv(tmpqubit, (rank+1)%size, 0, QMPI_COMM_WORLD);
					CNOT(qubits+num_spins-1, tmpqubit);
					Rz(tmpqubit, 2.0 * J * dt);
					CNOT(qubits+num_spins-1, tmpqubit);
					QMPI_Unrecv(tmpqubit, (rank+1)%size, 0, QMPI_COMM_WORLD);
					QMPI_Free_qmem(tmpqubit, 1);
				}
			}
		}
		for (unsigned site = 0; site < num_spins; ++site)
			Rx(qubits+site, -2.0*g*dt);
	}
}
int main() {
	QMPI_Init(0, 0);
	int rank, size;
	QMPI_Comm_size(QMPI_COMM_WORLD, &size);
	QMPI_Comm_rank(QMPI_COMM_WORLD, &rank);
	// Number of spins per node:
	unsigned num_local_spins = 2;
	// Number of annealing steps:
	double num_annealing_steps = 100;
	// Trotter number
	unsigned num_trotter = 1;
	double time = 1; // time to evolve per annealing step
	// Parameters of transverse-field Ising model
	double J = 0.; // coupling strength
	double g = 1.; // transverse field
	// allocate spins:
	auto qubits = QMPI_Alloc_qmem(num_local_spins);
	// init to ground state
	for (unsigned i = 0; i < num_local_spins; ++i)
		H(qubits+i);
	// run annealing schedule
	for (unsigned step = 0; step < num_annealing_steps; ++step) {
		J = step * 1.0/num_annealing_steps;
		g = 1.0-J;
		tfim_time_evolution(J, g, time, qubits, num_local_spins, num_trotter);
	}
	// Measure
	std::vector<int> res(num_local_spins);
	for (unsigned i = 0; i < num_local_spins; ++i)
		res[i] = Measure(qubits+i);
	QMPI_Free_qmem(qubits, num_local_spins);
	// Gather all (classical) results and output
	std::vector<int> allres(num_local_spins*size);
	MPI_Gather(&res[0], num_local_spins, MPI_INT, &allres[0], num_local_spins, MPI_INT, 0, QMPI_COMM_WORLD);
	if (rank == 0) {
		std::cout << "Measurements: ";
		for (auto r : allres)
			std::cout << r << " ";
		std::cout << std::endl;
	}
	QMPI_Finalize();
	return 0;
}
\end{lstlisting}

\end{document}